\documentclass[
    reprint,
    aps,
    superscriptaddress,
    amsmath,
    amssymb,
]{revtex4-2}

\usepackage{graphicx}
\usepackage{amsmath}
\usepackage{amssymb}
\usepackage{braket}

\usepackage{xcolor}

\usepackage[T1]{fontenc}
\usepackage[utf8]{inputenc}

\usepackage{microtype}

\usepackage{siunitx}

\DeclareSIUnit\dBm{dBm}
\DeclareSIUnit\dBc{dBc}

\makeatletter
\@ifclassloaded{revtex4-2}{%
    \setcounter{topnumber}{4}%
    \setcounter{totalnumber}{5}%
    \setcounter{dbltopnumber}{3}%
}{}
\makeatother

\newcommand{\qtyhy}[2]{\qty[quantity-product = \text{-}]{#1}{#2}}
\newcommand{\qtyrangehy}[3]{\num{#1}--\qtyhy{#2}{#3}}

\newcommand{\xCode}[1]{\texttt{#1}}

\newcommand{\xBackmatterSection}[1]{\par
        \vspace*{0.75\baselineskip}\section*{#1}}

\usepackage[colorlinks=true, linkcolor=blue!50!black, citecolor=blue!50!black, urlcolor=blue!50!black]{hyperref}

\hypersetup{
    pdfauthor={Akinori Machino et al.},
    pdftitle={QuBE/Qubex: an integrated hardware--software system for superconducting qubit experiments with broadband control}
}

\begin{document}

\title{QuBE/Qubex: an integrated hardware--software system\texorpdfstring{\\}{ }for superconducting qubit experiments with broadband control}
\newcommand{\xQiqbUOsaka}{Center for Quantum Information and Quantum Biology, The University of Osaka, Toyonaka, Osaka, Japan}
\newcommand{\xEsUOsaka}{Graduate School of Engineering Science, The University of Osaka, Toyonaka, Osaka, Japan}
\newcommand{\xQuel}{QuEL, Inc., Hachioji, Tokyo, Japan}
\newcommand{\xIstUOsaka}{Graduate School of Information Science and Technology, The University of Osaka, Suita, Osaka, Japan}
\newcommand{\xEtrees}{e-trees.Japan, Inc., Hachioji, Tokyo, Japan}
\newcommand{\xRiken}{RIKEN Center for Quantum Computing, Wako, Saitama, Japan}
\newcommand{\xEngUTokyo}{Graduate School of Engineering, The University of Tokyo, Bunkyo-ku, Tokyo, Japan}

\author{Akinori~Machino}
\email{akinori.machino@qiqb.osaka-u.ac.jp}
\affiliation{\xEsUOsaka}
\affiliation{\xQiqbUOsaka}
\affiliation{\xQuel}

\author{Kazuhisa~Ogawa}
\email{k-ogawa.qiqb@osaka-u.ac.jp}
\affiliation{\xQiqbUOsaka}

\author{Takefumi~Miyoshi}
\affiliation{\xQiqbUOsaka}
\affiliation{\xQuel}
\affiliation{\xEtrees}

\author{Hidehisa~Shiomi}
\affiliation{\xQiqbUOsaka}
\affiliation{\xQuel}

\author{Shinichi~Morisaka}
\affiliation{\xQiqbUOsaka}
\affiliation{\xQuel}

\author{Ryo~Matsuda}
\affiliation{\xEsUOsaka}
\affiliation{\xQiqbUOsaka}

\author{Nilton~F.~G.~Filho}
\affiliation{\xEsUOsaka}
\affiliation{\xQiqbUOsaka}

\author{Koichiro~Ban}
\affiliation{\xIstUOsaka}

\author{Takafumi~Miyanaga}
\affiliation{\xQiqbUOsaka}

\author{Keisuke~Koike}
\affiliation{\xEtrees}

\author{Ryutaro~Ohira}
\email{ohira@quel-inc.com}
\affiliation{\xQuel}

\author{Toshi~Sumida}
\email{sumida@quel-inc.com}
\affiliation{\xQuel}

\author{Yoshinori~Kurimoto}
\affiliation{\xQuel}

\author{Yuuya~Sugita}
\affiliation{\xQuel}

\author{Yosuke~Ito}
\affiliation{\xQuel}

\author{Yasunari~Suzuki}
\affiliation{\xRiken}
\affiliation{\xQiqbUOsaka}

\author{Peter~A.~Spring}
\affiliation{\xRiken}

\author{Shiyu~Wang}
\affiliation{\xRiken}

\author{Hiroto~Mukai}
\affiliation{\xEngUTokyo}
\affiliation{\xRiken}

\author{Arvind~Mamgain}
\affiliation{\xRiken}

\author{Shuhei~Tamate}
\affiliation{\xRiken}

\author{Yutaka~Tabuchi}
\affiliation{\xRiken}
\affiliation{\xQiqbUOsaka}

\author{Yasunobu~Nakamura}
\affiliation{\xRiken}
\affiliation{\xEngUTokyo}

\author{Makoto~Negoro}
\email{negoro.makoto.qiqb@osaka-u.ac.jp}
\affiliation{\xEsUOsaka}
\affiliation{\xQiqbUOsaka}
\affiliation{\xQuel}
\affiliation{\xIstUOsaka}

\date{\today}

\begin{abstract}
    Achieving high-fidelity operation in large-scale superconducting qubit systems requires not only control hardware with broad frequency coverage, low crosstalk, and tight synchronization but also software that coordinates system configuration, experiment execution, and data analysis.
    Here we present an integrated qubit-control system that combines broadband microwave hardware with a pulse-level software stack for scalable superconducting qubit experiments.
    The hardware provides broadband microwave coverage, including an instantaneous span of up to \(\qty{1.6}{\giga\hertz}\) from a control output, while the software reduces setup and calibration overhead through automated configuration and built-in experiment workflows.
    We validate the system on a 64-qubit fixed-frequency transmon chip through full-chip frequency identification and representative demonstrations, including multi-unit far-detuned cross-resonance calibration and benchmarking that yields a measured two-qubit gate fidelity of \(98.34\%\), and multilevel readout beyond the computational subspace.
    By disclosing the hardware architecture and releasing the software stack as open source, this work provides an inspectable hardware--software foundation for scalable superconducting qubit control experiments.
\end{abstract}

\maketitle

\section{Introduction}
\label{sec:introduction}

Operating large-scale superconducting qubit systems requires coordinated microwave generation, synchronization, readout, and calibration across distributed control hardware~\cite{krantz2019quantum, kjaergaard2020superconducting, motzoi2009simple, rizvi2026survey}. As superconducting quantum processors continue to scale, maintaining consistent configuration and timing across these functions becomes a system-level control problem.

Fixed-frequency transmon platforms are attractive for larger superconducting qubit systems because they avoid the flux-control lines required for tunable qubits while supporting all-microwave cross-resonance~(CR) gates~\cite{chow2011simple}. A key challenge, however, is that fabrication variability makes qubit-frequency allocation increasingly susceptible to frequency collisions as chip size grows~\cite{morvan2022frequency, hertzberg2021laser}. The far-detuned regime mitigates these constraints through qubit-frequency allocations with large detunings~\cite{ogawa2026highyield, inoue2026systematic}. This operating regime imposes a hardware requirement: the control hardware must provide broadband and high-power microwave control to support resonant single-qubit gates and CR drives between far-detuned qubits.

Broadband operation is also needed to access higher transmon levels relevant to leakage suppression, reset protocols, and qutrit-style operations~\cite{motzoi2009simple, magnard2018fast, liu2023performing}. The control-and-readout system must therefore support both broadband coherent control and readout of non-computational states such as the second excited state $\ket{f}$.

Even with suitable hardware, operating a large-scale qubit chip remains difficult when instrument settings across multiple control units require manual adjustments to match the qubit and readout-resonator frequency layout. Calibration state also has to be tracked consistently across experiments. Practical operation therefore requires software that resolves chip, wiring, and calibration information into executable instrument settings, so that the same setup and calibration procedures can be reproduced reliably on a given experimental system.

Recent open quantum-control systems and software frameworks have increasingly improved the inspectability and programmability of quantum-control stacks~\cite{stefanazzi2022qick, xu2021qubic, park2022icarus, efthymiou2024qibolab}. However, scalable far-detuned operation also requires an integrated demonstration in which broadband, high-power, synchronized control hardware is connected to software-driven configuration and calibration workflows. The system presented here addresses this gap by combining QuBE, a broadband microwave control and readout hardware platform, with Qubex, a pulse-level software framework.

QuBE, named for \emph{Qubit-controller with Broadband Electronics}, provides multi-unit microwave generation, readout capture, and synchronized clock distribution. Qubex provides automated system configuration, pulse-level experiment execution, and procedures for characterization, calibration, benchmarking, and measurement analysis. We disclose the hardware architecture in detail and release the software stack as open source.

We validate the system through representative workflows on a 64-qubit fixed-frequency transmon chip designed at RIKEN~\cite{tamate2022packaging, spring2025fast} and operated with the implemented control stack at The University of Osaka, including automated system setup, synchronized multi-unit control, broadband far-detuned CR calibration, and multilevel readout assisted by a Josephson parametric amplifier~(JPA). Because the chip design and fabrication are not the primary subject of this work, we treat the 64-qubit chip as the experimental platform and focus on the control system used to configure, operate, and validate it.

The remainder of this paper is organized as follows. Section~\ref{sec:architecture} describes the system architecture and the mapping from chip and system information to executable control settings. Section~\ref{sec:hardware-characterization} summarizes the relevant hardware characteristics of the QuBE platform. Section~\ref{sec:experiments} presents the experimental demonstrations on the superconducting qubit platform. Section~\ref{sec:discussion} discusses implications for larger-scale operation, and Section~\ref{sec:conclusion} concludes the paper.
\section{Control system architecture}
\label{sec:architecture}

This section describes the control-system architecture developed in this work. It focuses on how the hardware and software layers divide responsibility while remaining connected through a configuration and execution flow for scalable superconducting qubit experiments.

\subsection{System overview}

\begin{figure}[t]
    \centering
    \includegraphics[width=\columnwidth]{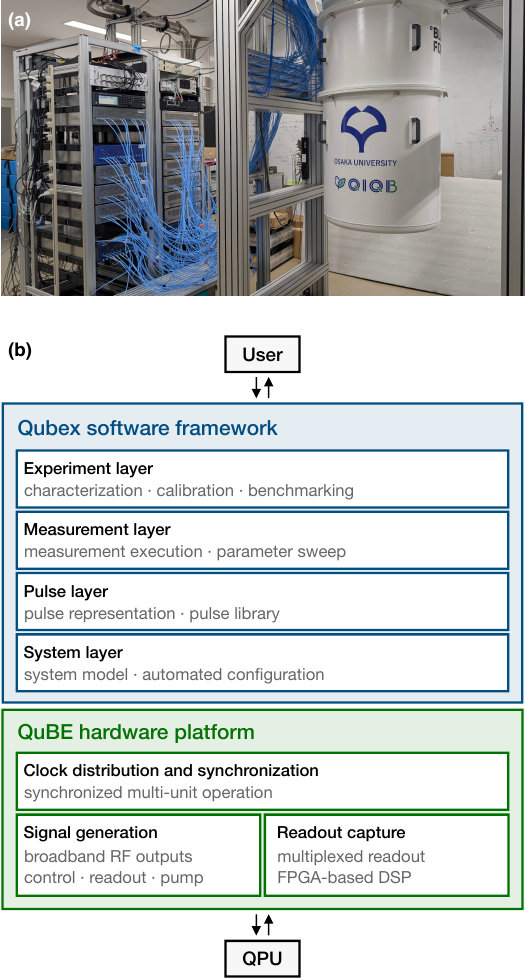}
    \caption{
        \label{fig:system-overview} System overview. (a) Photograph of the implemented QuBE control system. (b) Schematic of the hardware--software control stack used in this work. The Qubex software framework maps user-facing workflows through its experiment, measurement, pulse, and system layers, while the QuBE hardware platform provides clock distribution and synchronization, microwave signal generation, and readout capture for quantum processing unit (QPU) operation.
    }
\end{figure}

The platform integrates the QuBE hardware system and the Qubex software framework for large-scale superconducting qubit experiments. QuBE provides broadband microwave generation, readout capture, and synchronized clock distribution across multiple control units. In the 64-qubit configuration studied here, 12 QuBE units operate under shared clock and timing references and supply control and readout signals together with pump tones for parametric-amplifier readout. Fig.~\ref{fig:system-overview}(a) shows the implemented rack-scale system.

Qubex provides the pulse-level software environment for experiment configuration, execution, and analysis. It loads chip metadata, hardware inventory, wiring information, and calibration parameters, and exposes the resulting setup through modular experiment routines for characterization, calibration, benchmarking, and measurement analysis. Fig.~\ref{fig:system-overview}(b) summarizes the corresponding hardware--software stack, in which QuBE and Qubex together form an integrated control environment for scalable superconducting qubit experiments.

\subsection{Hardware architecture}

\begin{figure}[t]
    \centering
    \includegraphics[width=\columnwidth]{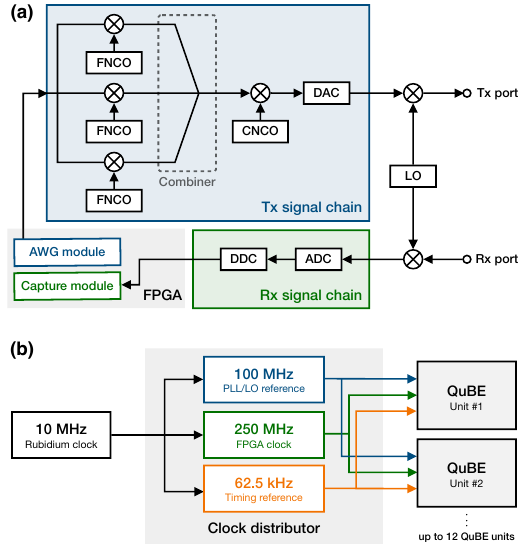}
    \caption{
        \label{fig:hardware-architecture} Hardware-side organization of the QuBE platform. (a) Representative transmit~(Tx) and receive~(Rx) signal chains within one QuBE unit. The Tx chain combines waveform generation implemented in the FPGA, FNCO/CNCO frequency shifts, DAC conversion, and shared-LO upconversion, while the Rx chain performs shared-LO downconversion, ADC digitization, DDC processing, and capture in the FPGA. The RF DAC/ADC signal chains are implemented using AD9082 mixed-signal converters. (b) Clock-distribution scheme for synchronized multi-unit operation. A rubidium-based \qtyhy{10}{\mega\hertz} clock is distributed as \qtyhy{100}{\mega\hertz} PLL/LO references, \qtyhy{250}{\mega\hertz} FPGA clocks, and \qtyhy{62.5}{\kilo\hertz} timing references for up to 12 QuBE units.
    }
\end{figure}

Across the QuBE platform, the hardware architecture integrates digital waveform generation, frequency conversion, readout capture, and internal monitoring paths in a compact mixed-signal design. Fig.~\ref{fig:hardware-architecture}(a) summarizes the transmit and receive signal paths. The control outputs are designed for qubit-drive frequencies in the \qtyrangehy{7.25}{10}{\giga\hertz} range, the readout outputs and inputs cover the \qtyrangehy{9.5}{11}{\giga\hertz} band, and dedicated parametric-amplifier pump outputs extend to the \qtyrangehy{19}{21}{\giga\hertz} range through a frequency-doubling stage. This frequency allocation allows the control band to support both single- and two-qubit gates, while the pump band supports JPA readout.

Control signals are constructed from digitally defined waveform components in the arbitrary waveform generator~(AWG) module implemented in a field-programmable gate array~(FPGA). In the transmit signal chain, outputs from multiple AWG units in this module can be frequency shifted by fine numerically controlled oscillator~(FNCO) paths before they are combined and shifted by a coarse numerically controlled oscillator~(CNCO). The combined signal is then converted by digital-to-analog converter~(DAC) resources and translated to the target microwave band through phase-locked-loop~(PLL)-based local oscillators~(LOs) and analog mixers.

The control ports used to drive CR tones support an instantaneous frequency span of up to \qty{1.6}{\giga\hertz} by combining three digitally generated waveform components shifted by separate FNCO paths; port-level implementation details are described in Appendix~\ref{sec:appendix-hardware-details}. This bandwidth allows the same physical control line to carry resonant single-qubit pulses and far-detuned CR tones without reconfiguring the analog signal path.

The receive signal chain performs readout and monitor capture. A returned readout or monitor signal is downconverted by an analog mixer driven by the shared LO, digitized by analog-to-digital converter~(ADC) resources, and passed through a digital downconverter~(DDC) before reaching the capture module implemented in the FPGA. The capture module performs readout-side digital signal processing~(DSP), including multiplexed demodulation, windowed integration, and state classification, yielding measurement data for higher-level software analysis. The QuBE unit also includes internal monitoring paths and switching circuitry for signal verification and timing checks without disturbing the main experimental wiring.

Multi-unit superconducting qubit control requires phase-coherent and time-aligned operation across distributed microwave channels. Fig.~\ref{fig:hardware-architecture}(b) shows the synchronization architecture used for this purpose. A dedicated clock distributor supplies shared \qtyhy{100}{\mega\hertz}, \qtyhy{250}{\mega\hertz}, and \qtyhy{62.5}{\kilo\hertz} reference signals to the units. The \qtyhy{100}{\mega\hertz} signal provides the reference for PLL-based local-oscillator generation, the \qtyhy{250}{\mega\hertz} signal drives the unit-side FPGA, and the \qtyhy{62.5}{\kilo\hertz} signal provides the timing reference used to synchronize experiment triggering. These references are distributed in a star topology from a central distribution stage. In the 64-qubit configuration demonstrated here, one distributor coordinates up to 12 QuBE units directly. Additional board-level implementation details of the QuBE signal chain, monitoring path, and receive-side DSP are summarized in Appendix~\ref{sec:appendix-hardware-details}.

\subsection{Software architecture}

The software stack is organized so that lower-level hardware-control software handles hardware-specific execution, while Qubex provides the experiment-facing framework above it. Within Qubex, responsibilities are separated into system, pulse, measurement, and experiment layers.

The system layer abstracts the qubit chip and the control hardware as software models and assembles a setup-specific experiment model from chip metadata, wiring information, and hardware parameters. This resolved experiment model provides a common configuration interface for the rest of the software stack.

The pulse layer provides a hardware-independent pulse-sequence representation shared across experiment workflows and execution targets. By separating pulse construction from backend-specific conversion, Qubex allows the same pulse-construction code to be translated either into QuBE register settings for hardware execution or into simulator-specific objects for numerical studies.

The measurement layer constructs pulse schedules, capture placement, parameter sweeps, and execution flows for pulse-level experiments using the shared pulse representation and resolved experiment model. Hardware-specific execution is delegated to the lower-level control software beneath Qubex, so the same workflow layer can in principle be retargeted to other compatible hardware backends.

The experiment layer provides built-in workflows for characterization, calibration, benchmarking, and related procedures. Users interact with experiment-oriented workflows rather than hardware-control commands directly. Additional notes on the whole software stack, including lower-level software components and their relation to Qubex, are summarized in Appendix~\ref{sec:appendix-software}.

\subsection{System integration}

\begin{figure}[t]
    \centering
    \includegraphics[width=0.95\columnwidth]{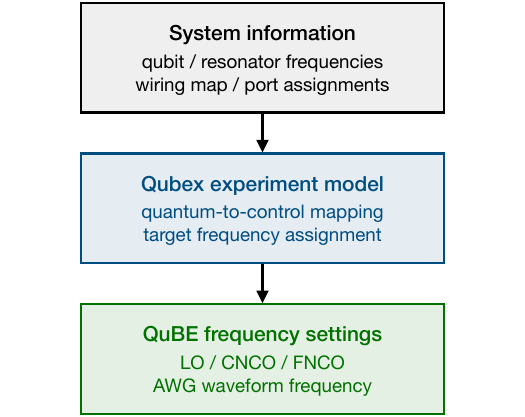}
    \caption{\label{fig:hardware-setting-resolution} Hardware-setting resolution in Qubex. System information, including qubit and resonator frequencies, wiring maps, and port assignments, is resolved into an experiment model that maps the quantum system to the control system. Target frequencies are then resolved into QuBE frequency settings consisting of LO, CNCO, FNCO, and AWG waveform-frequency terms.}
\end{figure}

Qubex resolves system information into executable QuBE settings through the flow summarized in Fig.~\ref{fig:hardware-setting-resolution}. A system definition, comprising chip metadata, hardware inventory, wiring information, and system-specific parameters, is first assembled into an experiment model that maps quantum-system objects such as qubits and resonators to control-system resources such as units, ports, and channels. Control and readout frequencies are then assigned on this model and decomposed into executable hardware frequency settings.

In this decomposition, each target frequency is expressed by LO, CNCO, FNCO, and a residual detuning in the waveform. In the configuration used here, control-band tones in the \qtyrangehy{7.25}{10}{\giga\hertz} range are generated from the lower sideband, whereas readout-band tones in the \qtyrangehy{9.5}{11}{\giga\hertz} range are generated from the upper sideband. The corresponding frequency relation is
\begin{equation}
    \label{eq:hardware-frequency-decomposition}
    F_{\mathrm{target}}
    =
    F_{\mathrm{LO}}
    +
    s
    \left(
    F_{\mathrm{CNCO}}
    +
    F_{\mathrm{FNCO}}
    +
    F_{\mathrm{AWG}}
    \right),
\end{equation}
where \(s=-1\) for lower-sideband control outputs and \(s=+1\) for upper-sideband readout outputs. Because the LO and NCO frequencies are constrained to discrete hardware-supported grids, the remaining offset is absorbed by the waveform-side term \(F_{\mathrm{AWG}}\). Qubex chooses these values so as to preserve the phase reproducibility required for coherent quantum control and to avoid unwanted aliasing or image tones in the generated microwave output; the grid constraints and the resolution strategy are detailed in Appendix~\ref{sec:appendix-frequency-settings}.

This resolution step separates experiment procedures from setup-specific configuration: experiment workflows are written against the experiment model and can be deployed on a particular setup without rewriting setup-specific logic.
\section{Hardware characterization}
\label{sec:hardware-characterization}

This section summarizes the hardware properties of the QuBE control platform that support the experiments in Sec.~\ref{sec:experiments}. For microwave-implemented superconducting qubit control and readout systems, relevant hardware requirements include frequency coverage, pulse-shape fidelity, suppression of spurious tones and crosstalk, stable phase references, timing alignment, and readout-chain performance~\cite{rizvi2026survey}. We therefore focus on broadband output capability, signal integrity and channel isolation, and phase stability and timing alignment.

\begin{figure}[t]
    \centering
    \includegraphics[width=1.0\columnwidth]{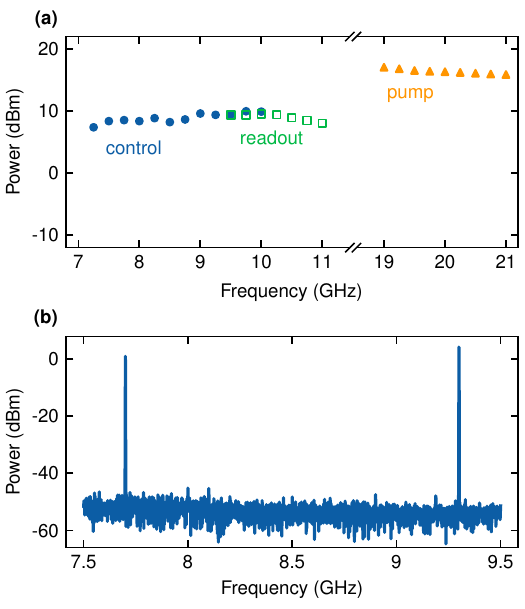}
    \caption{\label{fig:broadband-output} Broadband output capability of the QuBE controller. (a) Output power from representative control, readout, and pump ports across their operating bands. (b) Multi-tone output spectrum from one control output port, where two digital channels synthesize tones separated by \qty{1.6}{\giga\hertz} at the same physical output.}
\end{figure}

\begin{figure}[t]
    \centering
    \includegraphics[width=1.0\columnwidth]{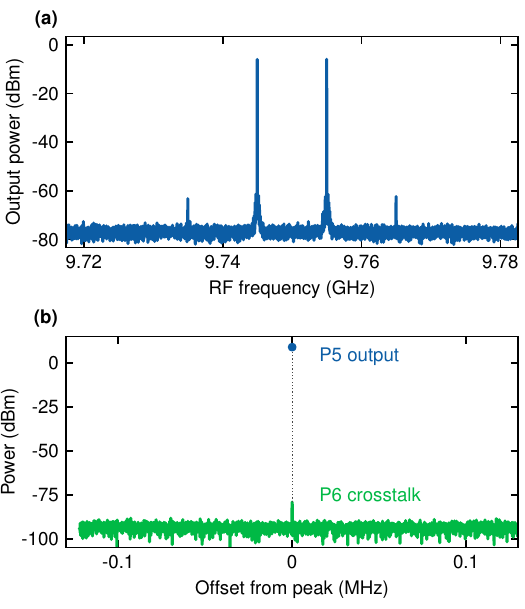}
    \caption{\label{fig:signal-integrity} Signal integrity and channel isolation of the QuBE controller. (a) Two-tone IMD measurement for a control channel. The programmed tones are placed around \qty{9.75}{\giga\hertz} with offsets of \qty{5}{\mega\hertz} and calibrated to approximately \qty{-6}{\dBm} per tone. The smaller side peaks are the third-order IMD products, which are at least \qty{56.4}{\decibel} below the driven tones. (b) Neighboring-port crosstalk measurement between driven P5 and adjacent P6 output ports. The P6 crosstalk peak is \qty{88.0}{\decibel} below the P5 target-port output.}
\end{figure}

\begin{figure}[t]
    \centering
    \includegraphics[width=1.0\columnwidth]{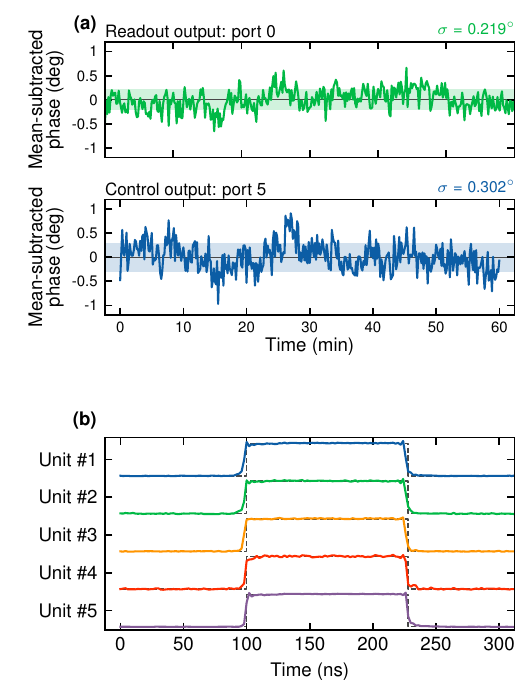}
    \caption{\label{fig:synchronization-stability} Phase stability and timing alignment of the QuBE controller. (a) Representative phase-stability traces measured at \qty{10}{\giga\hertz} from readout output port 0 and control output port 5 on the same QuBE unit. Both traces were acquired through monitor input port 4 over \qty{3600}{\second}. The phases were unwrapped and mean-subtracted. Plotted traces are sampled every \qty{10}{\second} for visibility, and shaded regions indicate \(\pm 1\) standard deviation of each mean-subtracted phase trace. (b) Multi-unit alignment measurement. Rectangular pulses of duration \qty{128}{\nano\second} are launched at $t=\qty{100}{\nano\second}$ from output ports on multiple QuBE units and acquired through a monitor port on a single unit under shared clock and trigger conditions. Colored traces show measured monitor waveforms, and dashed gray steps show the rectangular timing models used to identify pulse positions.}
\end{figure}

\subsection{Broadband output capability}

The hardware must provide the frequency coverage and output level needed across the control, readout, and pump paths. These paths are designed for qubit-drive tones in the \qtyrangehy{7.25}{10.0}{\giga\hertz} range, multiplexed readout tones in the \qtyrangehy{9.5}{11.0}{\giga\hertz} range, and parametric-amplifier pump tones in the \qtyrangehy{19.0}{21.0}{\giga\hertz} range, respectively. Fig.~\ref{fig:broadband-output}(a) shows representative output-power measurements across these operating bands. Because the effective drive strength at the qubit depends on the specific experimental setup, including cryogenic attenuation and chip-side control-line coupling, the controller-output power level alone does not define whether far-detuned CR operation is possible. We therefore treat Fig.~\ref{fig:broadband-output}(a) as hardware characterization and evaluate far-detuned CR operation experimentally in Sec.~\ref{sec:cross-resonance}.

To drive CR gates in a far-detuned regime where neighboring qubits can be separated by about \qty{1}{\giga\hertz}, the control path supports an instantaneous frequency span of up to \qty{1.6}{\giga\hertz}. This capability allows one control line to synthesize both resonant single-qubit pulses and CR drives between far-detuned qubits. Fig.~\ref{fig:broadband-output}(b) shows representative multi-tone synthesis from two digital channels at one control output port. The readout and pump paths likewise provide the microwave coverage required for the JPA-assisted multilevel readout experiment described in Sec.~\ref{sec:multilevel-readout}.

\subsection{Signal integrity and channel isolation}

Broadband signal generation must maintain spectral purity under experimental operating conditions. Fig.~\ref{fig:signal-integrity}(a) shows a representative two-tone measurement used to evaluate intermodulation distortion~(IMD). The two tones are generated around a \qty{9.75}{\giga\hertz} center frequency with offsets of \qty{5}{\mega\hertz} and calibrated to approximately \qty{-6}{\dBm} per tone. In the representative spectrum shown here, the third-order IMD products are at least \qty{56.4}{\decibel} below the driven tones. This representative IMD suppression supports pulse-shaping workflows that require clean programmed spectra, including selective excitation on shared control lines~\cite{matsuda2026selective}.

Channel isolation is likewise important because the QuBE hardware integrates many microwave channels within a compact form factor. Fig.~\ref{fig:signal-integrity}(b) shows a neighboring-port crosstalk measurement. The P5 target-port output is measured near \qty{9.75}{\giga\hertz}, giving an absolute output level of \qty{8.86}{\dBm}. With this output maintained, the adjacent P6 output level is measured. The P6 crosstalk peak is \qty{88.0}{\decibel} below the driven P5 output for the representative adjacent-port pair shown here. This isolation helps keep simultaneous multiqubit control from being dominated by controller-side crosstalk.

\subsection{Phase stability and timing alignment}

For coherent qubit control, the relevant hardware requirement is stable microwave phase over the timescales on which calibrations and repeated measurements are executed. We evaluated the output phase of a QuBE unit by measuring \qty{10}{\giga\hertz} continuous-wave signals through monitor paths. Fig.~\ref{fig:synchronization-stability}(a) shows representative \qty{3600}{\second} traces from a readout output and a control output after phase unwrapping and mean subtraction. The corresponding standard deviations are \qty{0.219}{\degree} and \qty{0.302}{\degree}, respectively. Applying the same analysis to the six measured output ports acquired through monitor paths gives standard deviations below \qty{0.456}{\degree}, indicating that the representative traces are consistent with the port-to-port behavior measured on the unit. A related study on QuEL-1 SE, a controller in the same hardware family, demonstrates \qty{24}{\hour} microwave-output stabilization using device-level temperature control of analog components~\cite{kurimoto2026stabilization}.

Timing alignment across units is likewise required for distributed pulse execution. This requirement is particularly important for multi-unit CR control with active cancellation: the CR drive and cancellation tone are emitted from ports on different control units, and their relative timing must be tightly aligned at the chip. In the skew-calibration workflow, rectangular test pulses from target output ports are acquired through a common monitor path and converted into pulse-start estimates based on a fixed rectangular window of 64 samples, corresponding to \qty{128}{\nano\second}, over the measured magnitude trace. The resulting indices are then used to update per-unit delay settings. In the representative alignment shown in Fig.~\ref{fig:synchronization-stability}(b), the fitted pulse starts for the selected ports coincide at $t=\qty{100}{\nano\second}$, corresponding to the \qtyhy{2}{\nano\second} sampling grid of the monitor trace.
\section{Experimental demonstrations}
\label{sec:experiments}

This section demonstrates the integrated control system on a 64-qubit superconducting qubit platform. The demonstrations cover automated setup from chip and hardware metadata, single-qubit characterization and gate validation, broadband multi-unit far-detuned CR calibration, and JPA-assisted multilevel readout of the $\ket{g}$, $\ket{e}$, and $\ket{f}$ transmon states.

\subsection{Experimental setup}
\label{sec:experimental-setup}

The experiments were performed on a 64-qubit fixed-frequency transmon chip developed at RIKEN within a scalable packaging architecture~\cite{tamate2022packaging, spring2025fast} and operated with the QuBE/Qubex control stack installed at The University of Osaka. The chip is cooled to millikelvin temperatures in a dilution refrigerator to operate as a superconducting circuit, suppress thermal excitation, and preserve qubit coherence.

\begin{figure}[t]
    \centering
    \includegraphics[width=1.0\columnwidth,trim={0 52bp 0 0},clip]{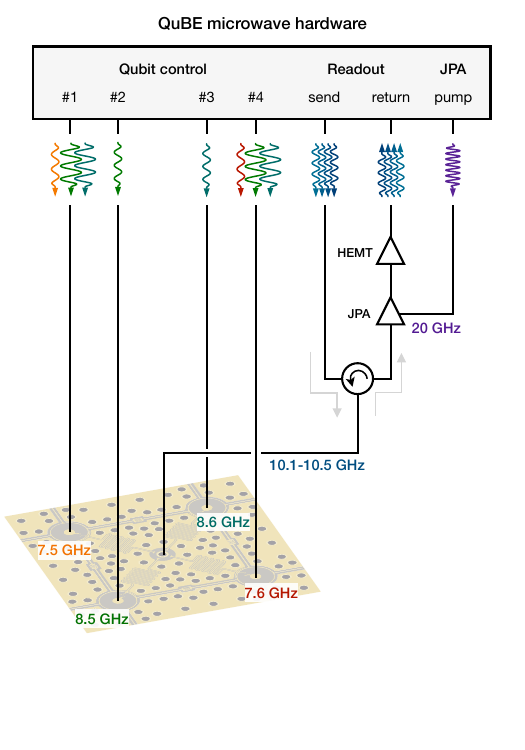}
    \caption{\label{fig:experimental-wiring-overview} Wiring between the QuBE microwave hardware and the qubit chip for one four-qubit readout-multiplexing unit. Four QuBE control output ports drive dedicated qubit-control lines, while the readout-send, readout-return, and JPA-pump paths are shared within the unit. The readout-return path includes JPA and HEMT amplification before returning to the QuBE readout input. Representative frequencies indicate the qubit-drive tones, frequency-multiplexed readout tones, and JPA-pump tone.}
\end{figure}

The wiring between the QuBE microwave hardware and the qubit chip is grouped into four-qubit readout-multiplexing units, as shown in Fig.~\ref{fig:experimental-wiring-overview}. In each unit, four dedicated control lines drive the individual qubits, whereas the readout-send, readout-return, and JPA-pump lines are shared among the four qubits. Extending this unit layout to the 64-qubit chip gives 64 control lines and 16 lines of each shared type. The corresponding cryostat wiring, including staged attenuation, filtering, and temperature-stage placement, is described in Appendix~\ref{sec:appendix-cryostat-wiring}.

For readout, the readout-send line delivers frequency-multiplexed tones to the readout resonators, and the readout-return line carries the reflected signals back to the QuBE readout input ports for digitization and FPGA-based demodulation. On the readout-return path, a JPA serves as a low-noise preamplifier for superconducting qubit readout~\cite{yamamoto2008flux,lin2013single}. In our setup, this stage uses a RIKEN-designed impedance-engineered parametric amplifier~(ImPA) driven through the JPA-pump line and followed by a high-electron-mobility transistor~(HEMT) amplifier. The control lines carry both single-qubit drives and far-detuned CR drives, using the broadband output capability described in Sec.~\ref{sec:hardware-characterization}.

\begin{figure*}[t]
    \centering
    \includegraphics[width=1.0\textwidth]{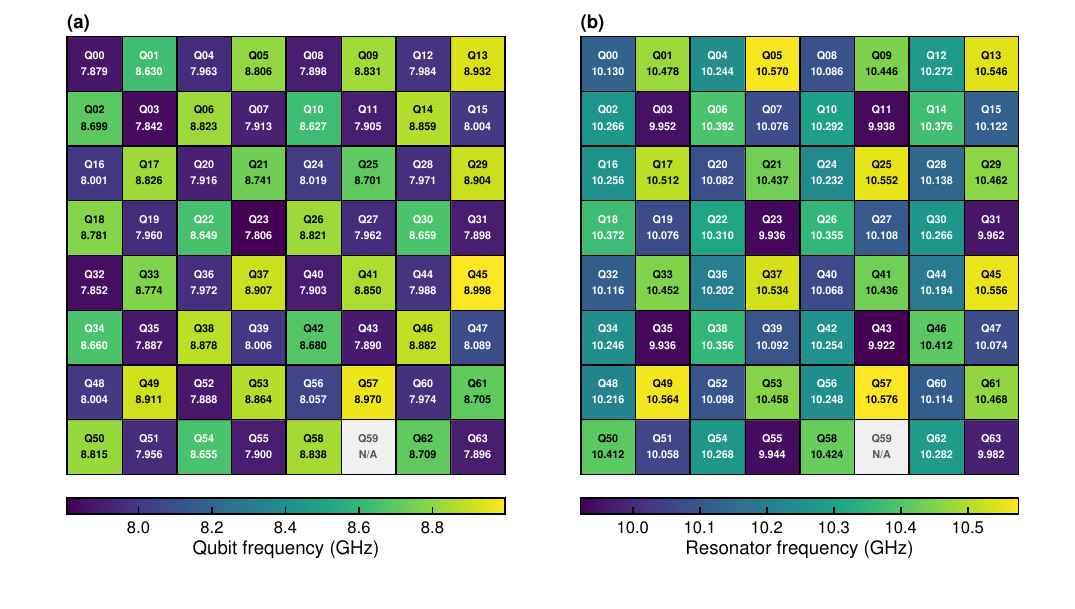}
    \caption{\label{fig:chip-frequency-map} Frequency allocation of the 64-qubit fixed-frequency transmon chip, measured using the QuBE/Qubex control stack. (a) Qubit frequencies, showing a checkerboard-like allocation of lower- and higher-frequency qubits with nearest-neighbor frequency differences of about \qty{1}{\giga\hertz}. (b) Readout-resonator frequencies. Cell labels follow the Qubex qubit identifiers, and thick grid lines indicate the four-qubit readout-multiplexing grouping used by the control-system configuration. Q59 is a defect site for which no resonator resonance was confirmed.}
\end{figure*}

Before each experiment, Qubex loads chip metadata, hardware inventory, wiring information, and calibration state from a system definition and resolves them into a concrete experimental setup. The frequency information summarized in Fig.~\ref{fig:chip-frequency-map} is one input to this setup procedure. Qubex then resolves these inputs into the frequency settings for the hardware as described in Sec.~\ref{sec:architecture}~(Fig.~\ref{fig:hardware-setting-resolution}). A representative example is given in Table~\ref{tab:control-frequency-settings} in Appendix~\ref{sec:appendix-frequency-settings}.

Table~\ref{tab:device-parameters} lists physical parameters of qubits highlighted in the focused demonstrations below. Full-chip qubit and resonator frequencies are shown in Fig.~\ref{fig:chip-frequency-map}.

\begin{table*}[t]
    \caption{\label{tab:device-parameters} Physical parameters of qubits highlighted in the focused demonstrations of Sec.~\ref{sec:experiments}.}
    \begin{ruledtabular}
        \begin{tabular*}{\textwidth}{@{\extracolsep{\fill}}lcccccc}
            Label      & Role                              & $f_r$ (\si{\giga\hertz}) & $f_q$ (\si{\giga\hertz}) & $\alpha$ (\si{\mega\hertz}) & $T_1$ (\si{\micro\second}) & $T_{2,\mathrm{echo}}$ (\si{\micro\second}) \\
            \colrule
            Q10        & multilevel readout                & 10.304                   & 8.627                    & -436.3                      & 37.2                       & 55.4                       \\
            Q22        & 1Q RB                             & 10.285                   & 8.649                    & -441.3                      & 33.2                       & 13.9                       \\
            Q25        & CR target                         & 10.547                   & 8.701                    & -475.3                      & 21.2                       & 24.9                       \\
            Q28        & coherence; CR control             & 10.127                   & 7.971                    & -362.9                      & 58.9                       & 52.1                       \\
            Q37        & frequency identification          & 10.516                   & 8.907                    & -468.7                      & 31.8                       & 12.2                       \\
        \end{tabular*}
    \end{ruledtabular}
\end{table*}

\subsection{Single-qubit workflows}
\label{sec:single-qubit-workflows}

\begin{figure}[t]
    \centering
    \includegraphics[width=1.0\columnwidth]{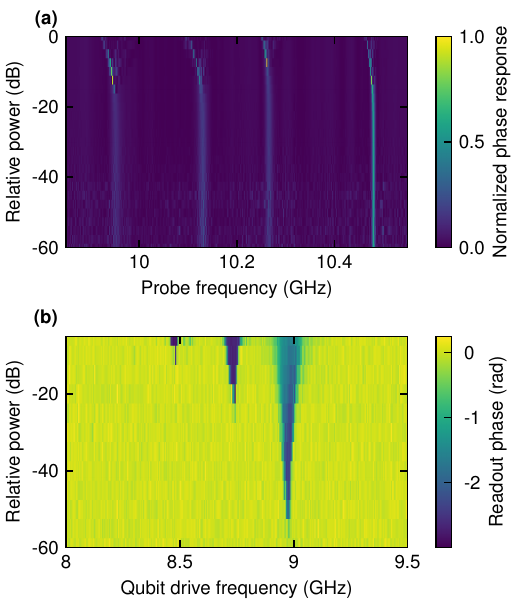}
    \caption{\label{fig:frequency-identification} Frequency-identification measurements. (a) Readout-resonator spectroscopy for the Q00--Q03 readout-multiplexing unit, plotted as a normalized phase-response map versus probe frequency and relative probe power. The four resonator responses appear as bright ridges at the sharp phase-change points, and at high probe power the ridges shift toward the bare resonator frequencies. (b) Power-dependent qubit spectroscopy for Q37, plotted as the readout phase versus qubit-drive frequency and relative drive power. The computational $g$--$e$ transition appears as a dominant feature, while increasing the drive power reveals the two-photon $g$--$f$ response and the $e$--$f$ transition.}
\end{figure}

Initial frequency characterization is performed through readout-resonator and qubit spectroscopy. These measurements are standard initial steps in microwave-control calibration workflows~\cite{rizvi2026survey} and are executed directly within the QuBE control system without using a separate vector network analyzer or other external instruments. In the resonator scan, Qubex sweeps a probe tone on the readout line; at each probe point, it applies a flat-top readout pulse that is long compared with the resonator ring-up time and integrates the captured response over the pulse window, so that the pulsed sweep approximates continuous-wave spectroscopy. Qubex measures the averaged complex readout response, corrects the electrical-delay phase, and identifies resonator frequencies from sharp phase changes in the corrected response. Wide scans are managed by stepping the shared LO and CNCO settings between subranges and sweeping the waveform-side frequency within each subrange; the subranges are stitched together in software by remeasuring the boundary point to remove inter-subrange phase offsets, yielding a continuous phase response across the full scan range. In the qubit scan, Qubex sweeps the qubit-drive frequency over multiple drive-power settings and monitors the readout-phase response at the selected resonator frequency. Fig.~\ref{fig:frequency-identification} demonstrates these frequency-identification workflows with a readout-resonator phase-response map for one readout-multiplexing unit and a power-dependent qubit-spectroscopy phase map.

\begin{figure}[t]
    \centering
    \includegraphics[width=1.0\columnwidth]{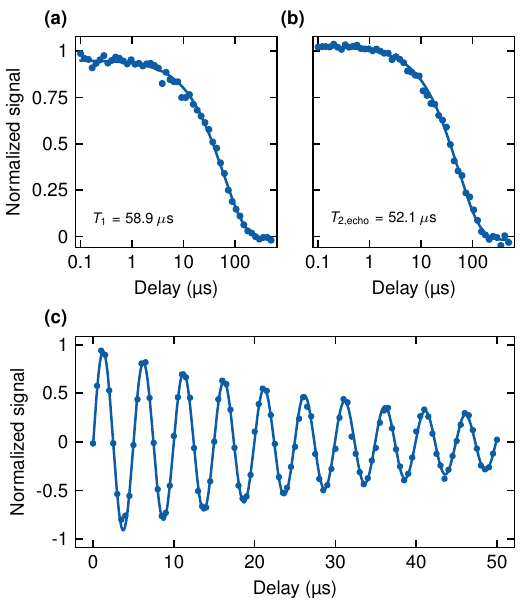}
    \caption{\label{fig:coherence-summary} Coherence measurements for Q28. Markers show measured data, and solid lines show fitted model curves. (a) $T_1$ relaxation measurement. (b) $T_{2,\mathrm{echo}}$ measurement. (c) Ramsey fringe measured with a drive detuning of \(\qty{0.2}{\mega\hertz}\). The fitted oscillation is used to refine the qubit resonance frequency.}
\end{figure}

Fig.~\ref{fig:coherence-summary}(a)--(c) reports $T_1$, $T_{2,\mathrm{echo}}$, and Ramsey measurements for one of the highlighted qubits. These measurements characterize the qubit coherence properties and, in the case of the Ramsey experiment, provide a precise determination of the qubit resonance frequency.

\begin{figure}[t]
    \centering
    \includegraphics[width=1.0\columnwidth]{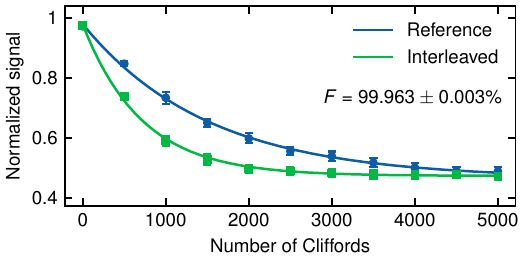}
    \caption{\label{fig:1q-rb} Single-qubit gate-validation measurement on the superconducting qubit platform. Interleaved randomized benchmarking~(IRB) results for a calibrated Q22 \(X_{\pi/2}\) gate are plotted together with the reference randomized benchmarking results. Each data point averages 30 random Clifford sequences at the corresponding sequence length. The error bars show the standard deviation across those sequences, and solid lines show exponential decay fits. The quoted fidelity uncertainty is the fit uncertainty from the IRB decay analysis.}
\end{figure}

For gate validation, Fig.~\ref{fig:1q-rb} gives an interleaved randomized benchmarking~(IRB)~\cite{magesan2012efficient} measurement for a calibrated \(X_{\pi/2}\) gate. The gate is implemented with a DRAG pulse, whose quadrature component is shaped to suppress leakage to higher transmon levels~\cite{motzoi2009simple, chow2010optimized, gambetta2011analytic}. In IRB, the decay of a reference Clifford sequence~\cite{knill2008randomized, magesan2011scalable} is compared with that of a sequence interleaving the target gate, providing a gate-specific estimate of the target-gate fidelity. The resulting \(X_{\pi/2}\) fidelity is \(99.963 \pm 0.003\%\).

These experiment workflows can be launched interactively by an experimenter or invoked programmatically for repeated execution. The full-chip frequency map in Fig.~\ref{fig:chip-frequency-map} was obtained by systematically running the Qubex spectroscopy procedures across the chip.

\subsection{Far-detuned cross-resonance calibration}
\label{sec:cross-resonance}

The CR gate is an all-microwave two-qubit gate in which driving the control qubit at the neighboring target-qubit frequency generates an effective interaction that makes the target-qubit rotation depend on the control-qubit state~\cite{chow2011simple}. Let \(f_{\mathrm{c}}\) and \(f_{\mathrm{t}}\) denote the control- and target-qubit frequencies, and let \(\Delta_{\mathrm{ct}}=f_{\mathrm{c}}-f_{\mathrm{t}}\). In a far-detuned regime, the magnitude of this control--target detuning exceeds the transmon anharmonicity scale~\cite{inoue2026systematic}. Consequently, the microwave output port assigned to the control qubit must synthesize two widely separated frequencies under the same output configuration, with resonant pulses at \(f_{\mathrm{c}}\) and CR pulses at \(f_{\mathrm{t}}\). This requirement is supported by the broadband output capability characterized in Sec.~\ref{sec:hardware-characterization}~(Fig.~\ref{fig:broadband-output}).

Gate calibration uses Hamiltonian tomography of the effective CR interaction by comparing target-qubit dynamics for the control qubit prepared in its \(\ket{g}\) and \(\ket{e}\) states. The extracted conditional response quantifies the desired \(ZX\) term, while an active cancellation tone is tuned to suppress unwanted target-qubit rotations from classical and quantum crosstalk~\cite{sheldon2016crosstalk}. Qubex implements these steps in a built-in calibration workflow.

The calibrated gate is refined using an echoed construction and a rotary component. The echoed CR sequence applies CR pulses with opposite signs around an echo pulse, canceling unwanted rotations that are unchanged by the CR-pulse sign reversal~\cite{sheldon2016crosstalk}. The rotary echo reduces residual coherent errors during the gate~\cite{sundaresan2020rotary}. Additional details on the effective-interaction picture and the calibration procedure are summarized in Appendix~\ref{sec:appendix-cr-details}.

Fig.~\ref{fig:cross-resonance}(a) illustrates the resulting multi-unit echoed CR sequence. We use the Q28--Q25 pair here, which spans multiple readout-multiplexing groups and is driven across multiple QuBE control units. The Q28--Q25 frequency placement is far-detuned, with \(\Delta_{\mathrm{ct}}=\qty{-729.284}{\mega\hertz}\); the detuning magnitude exceeds the magnitudes of both the target- and control-qubit anharmonicities, \(\alpha_{\mathrm{t}}=\qty{-475.3}{\mega\hertz}\) and \(\alpha_{\mathrm{c}}=\qty{-362.9}{\mega\hertz}\). In the sequence, the control-qubit output port emits the echo \(\pi\) pulses at \(f_{\mathrm{c}}\) and the CR pulses at \(f_{\mathrm{t}}\) under the same output configuration, while a separate unit applies the target-side cancellation pulse at \(f_{\mathrm{t}}\). Execution on the participating units is launched at a scheduled value of the synchronized time counter~(Appendix~\ref{sec:appendix-hardware-details}), and the distributed pulses are aligned using the skew-calibration workflow characterized in Sec.~\ref{sec:hardware-characterization} and Fig.~\ref{fig:synchronization-stability}(b); this workflow measures inter-unit skew and applies nanosecond-scale timing compensation before execution.

Fig.~\ref{fig:cross-resonance}(b) presents an interleaved randomized benchmarking result for the calibrated far-detuned CR gate on this pair. The extracted two-qubit gate fidelity is \(98.34 \pm 0.08\%\). This benchmark exercises the broadband, high-output-power, and synchronized microwave-control paths required for a multi-unit far-detuned CR workflow on this platform. The calibrated pulse parameters used for this benchmark are summarized in Appendix~\ref{sec:appendix-cr-details}.

\begin{figure}[t]
    \centering
    \includegraphics[width=1.0\columnwidth]{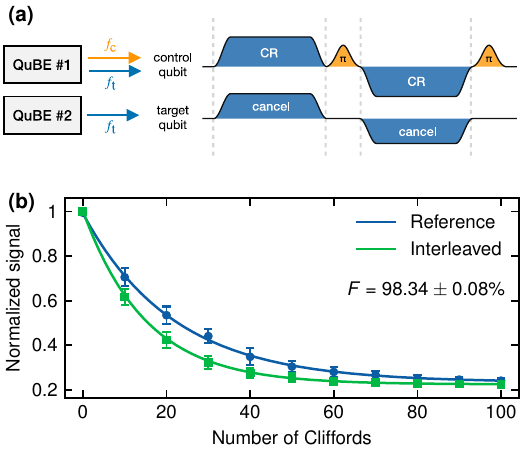}
    \caption{\label{fig:cross-resonance} Far-detuned cross-resonance calibration. (a) Schematic of multi-unit echoed CR pulse execution. QuBE~\#1 drives the control-qubit line with CR pulses at the target-qubit tone \(f_{\mathrm{t}}\) and echo \(\pi\) pulses at the control-qubit tone \(f_{\mathrm{c}}\), while QuBE~\#2 applies a target-side cancellation/rotary pulse at \(f_{\mathrm{t}}\). The cancellation component suppresses unwanted target rotations, and the rotary component intentionally adds the \(IX\)-like drive used during the echoed gate. Dashed vertical lines indicate pulse-boundary timing points that must remain aligned across multiple QuBE units. (b) Interleaved randomized benchmarking result for the calibrated Q28--Q25 far-detuned cross-resonance gate. Each data point averages 30 random Clifford sequences at the corresponding sequence length. The error bars show the standard deviation across those sequences. The quoted fidelity uncertainty is the fit uncertainty from the IRB decay analysis.}
\end{figure}

\subsection{Beyond-computational-subspace readout}
\label{sec:multilevel-readout}

Transmon excitation levels beyond the first two used as the computational subspace appear as leakage channels during pulse calibration and are also used deliberately in shelving readout, reset, and qutrit-style protocols~\cite{motzoi2009simple, chen2023transmon, magnard2018fast, liu2023performing}. The control stack therefore needs readout and analysis capabilities that can resolve the second excited state, $\ket{f}$, in addition to the computational $\ket{g}$ and $\ket{e}$ states.

\begin{figure}[t]
    \centering
    \includegraphics[width=1.0\columnwidth]{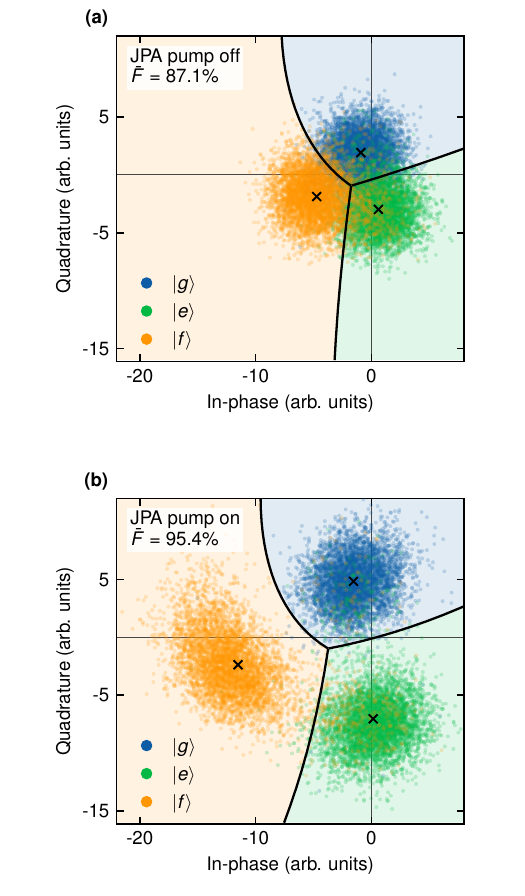}
    \caption{\label{fig:multilevel-readout} Beyond-computational-subspace readout. (a) Q10 multilevel readout classification in the I/Q plane with the JPA pump disabled. The circle markers show single-shot measurements prepared in the $\ket{g}$, $\ket{e}$, and $\ket{f}$ states, crosses show empirical cluster centers, and shaded regions with black curves indicate the Qubex Gaussian-mixture-model decision regions and boundaries. (b) Same readout measurement with the JPA pump enabled; the average state-assignment fidelity \(\bar{F}\) increases from \(87.1\%\) to \(95.4\%\).}
\end{figure}

We evaluate this capability through single-shot readout with the JPA pump disabled and enabled. For each condition, Qubex prepares the $\ket{g}$, $\ket{e}$, and $\ket{f}$ states and acquires integrated I/Q samples through the QuBE readout chain, whose FPGA-side DSP performs the demodulation and integration. State discrimination is then performed in software: Qubex classifies the resulting I/Q distributions with a Gaussian-mixture-model classifier, rather than using the hardware-level threshold classification also available in the capture module~(Appendix~\ref{sec:appendix-hardware-details}). With the JPA pump disabled, the average state-assignment fidelity is \(87.1\%\), as shown in Fig.~\ref{fig:multilevel-readout}(a). When the JPA pump is enabled, Fig.~\ref{fig:multilevel-readout}(b) shows improved cluster separation, and the average fidelity increases to \(95.4\%\). This average is the arithmetic mean of the correct-assignment probabilities for the prepared $\ket{g}$, $\ket{e}$, and $\ket{f}$ states, which are \(99.5\%\), \(95.7\%\), and \(91.2\%\), respectively, for the pump-enabled measurement. The detailed assignment matrix is given in Appendix~\ref{sec:appendix-multilevel-readout}.

This measurement validates the combined use of the QuBE readout and pump paths with Qubex-side classification for non-computational-state readout. The ability to distinguish $\ket{g}$, $\ket{e}$, and $\ket{f}$ supports leakage-aware calibration and protocols involving higher transmon levels.

\section{Discussion}
\label{sec:discussion}

In this work, we demonstrated automated system setup, full-chip frequency identification, multi-unit far-detuned CR calibration, and JPA-assisted multilevel readout on a 64-qubit superconducting qubit platform. In particular, the multi-unit far-detuned CR calibration shows that the integrated control stack provides the broadband frequency coverage, sufficient drive power, and synchronized qubit control required to operate a fixed-frequency, far-detuned chip architecture. The JPA-assisted multilevel-readout experiment further verifies readout and analysis beyond the computational subspace within the same environment.

At the system level, these results clarify the functional separation between QuBE and Qubex. QuBE supplies the hardware functions characterized in Sec.~\ref{sec:hardware-characterization}, including broadband output capability, signal integrity and channel isolation, and phase stability and timing alignment. Qubex maps chip and wiring information to executable settings and organizes setup, spectroscopy, calibration, and benchmarking as built-in pulse-level procedures. Together, these layers enable reproducible setup and pulse-level execution within one coordinated control environment.

The Qubex software abstraction also makes the pulse-level control stack suitable as a backend for higher-level calibration management and service orchestration. For example, connecting Qubex to service tools such as OQTOPUS/QDash provides a path for gate-level cloud jobs to be translated into calibrated pulse-level execution~\cite{kakuko2025practical, yoshida2025auxiliary, do2026dftembedded}. Agent-based calibration frameworks such as the NVIDIA Quantum Calibration Agent Blueprint~\cite{nvidia2026qcablueprint} illustrate a complementary direction toward autonomous orchestration.

At the same time, the present results remain representative validation of the implemented platform. Future extensions of this work include chip-wide calibration coverage, statistical evaluation across multiple CR pairs, long-term stability studies, and closed-loop autonomous calibration. On the hardware side, the clock distribution offers a concrete scaling path: the fanout stages can be cascaded hierarchically, extending the shared reference distribution well beyond the present 12-unit installation at the cost of additional phase noise introduced per stage. Larger systems will also remain constrained by hardware-infrastructure issues common to scalable microwave control and readout stacks, including wiring density, communication bandwidth, thermal load, and low-latency processing~\cite{rizvi2026survey}. These challenges indicate that scalable superconducting qubit operation will continue to require co-development of the qubit platform, the control hardware, and the software orchestration layer.
\section{Conclusion}
\label{sec:conclusion}

We presented QuBE and Qubex as an integrated control system for superconducting qubit experiments. The hardware characterization demonstrated broadband output capability, signal integrity, and synchronized multi-unit operation of the QuBE platform, while the experimental demonstrations validated automated setup, far-detuned CR calibration, and JPA-assisted multilevel readout on a 64-qubit superconducting qubit platform.

These results show that the implemented system supports the broadband-control workflows, synchronized operation, and reproducible pulse-level execution for scalable superconducting qubit experiments. By disclosing the hardware architecture and releasing the software stack as open source, this work provides an inspectable and reusable platform for future large-scale superconducting qubit control studies.

\appendix

\section{Hardware details}
\label{sec:appendix-hardware}
\label{sec:appendix-hardware-details}

\begin{figure}[t]
    \centering
    \includegraphics[width=0.95\columnwidth]{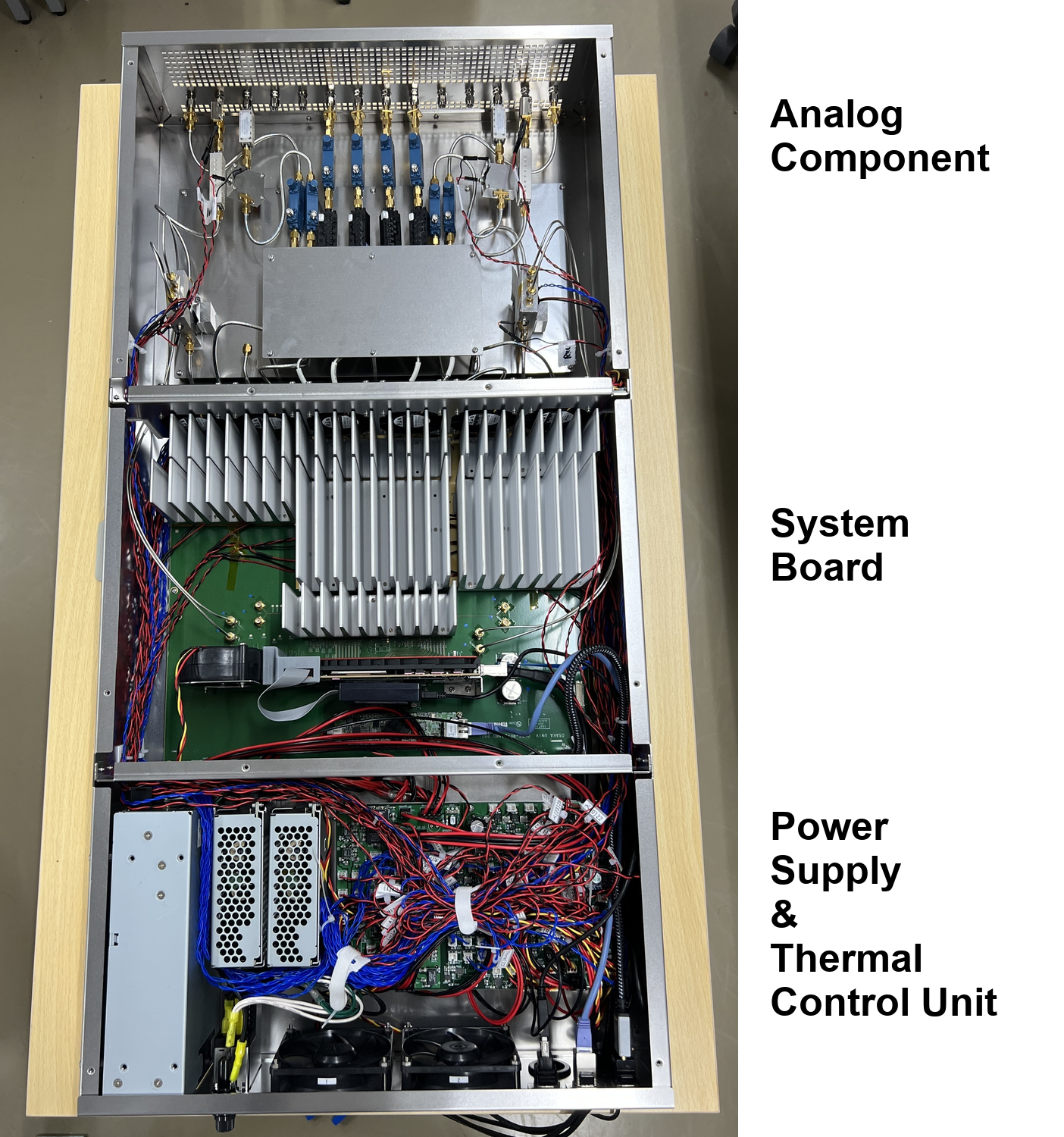}
    \caption{\label{fig:qube-unit-internal} Internal view of a QuBE unit. The upper region contains packaged analog microwave components such as filters, multiplexers, and combiners. The middle region contains the system board with the FPGA and mixed-signal resources, and the lower region contains the power-supply and thermal-control electronics.}
\end{figure}

At rack scale, the implemented QuBE (\emph{Qubit-controller with Broadband Electronics}) installation consists of 12 units arranged in a 19-inch format, each occupying 3U and together providing a control system for up to 64 qubits. Together, the 12 units fit within approximately a \(1~\mathrm{m} \times 1~\mathrm{m} \times 1~\mathrm{m}\) cube, and the total power consumption is approximately \qty{2.4}{\kilo\watt} during typical operation. The system comprises two unit types, Unit A and Unit B, each featuring 14 microwave ports. In Unit A these ports are assigned as four control ports, two readout output ports, two readout input ports, two pump output ports, two monitor output ports, and two monitor input ports. Unit B instead provides eight control ports, two no-connection positions, two monitor output ports, and two monitor input ports. Fig.~\ref{fig:qube-unit-internal} shows the internal layout of one QuBE unit.

The control-system connections are separated by function. Waveform transfer and capture-data readout use each unit's 10-GbE Wave interface, which is handled by the unit's Alveo U50 board. Register-level configuration uses a separate unit-side 1-GbE Config interface through a configuration hub on the isolated control-system network. For multi-unit operation, the host 10-GbE network interface card~(NIC) and the unit Wave interfaces are connected to a common Wave hub. The clock master connects to this Wave hub through one 10-GbE lane and to a separate Sync hub through another lane, so that waveform traffic and time-counter synchronization are carried on separate network paths. The Alveo U200 clock-master FPGA distributes counter information over this 10-GbE Sync path and is not the endpoint for waveform payload transfer. In this topology, the waveform and capture traffic of all units shares the host-side link, so the per-unit transfer bandwidth decreases as more units transfer data simultaneously. The communication bandwidth can be scaled by upgrading the host-side network interfaces and switches, and the FPGA-side data reduction through integration and averaging keeps the capture traffic well below the capacity of the shared link in typical calibration workflows. Waveform synthesis on the FPGA, in which pulses are compiled from parameters rather than transferred as sampled data, could further reduce the host-to-unit traffic. Reference-clock synchronization is provided separately through a dedicated clock-distribution link carrying shared \qtyhy{100}{\mega\hertz}, \qtyhy{250}{\mega\hertz}, and \qtyhy{62.5}{\kilo\hertz} signals. These references are generated by a clock distributor built around an AD9528 clock generator (Analog Devices, Inc.) together with ADCLK950 and ADCLK944 fanout buffers (Analog Devices, Inc.), and the distributor can also lock to an external \qtyhy{10}{\mega\hertz} reference.

Multi-unit experiment execution is started through the time-counter synchronization carried on the Sync path. The host software schedules each execution at a designated future value of the shared time counter, and every participating unit starts its waveform generation and capture when its local counter reaches the scheduled value. Because the counters are clocked by the shared references and aligned through the Sync path, distributed pulse sequences such as the multi-unit CR execution in Sec.~\ref{sec:cross-resonance} start with a deterministic timing relation across units.

Inside each unit, the central implementation element is a system board that accommodates two FPGA boards and the mixed-signal converter resources. An exStickGE board (e-trees.Japan, Inc.) is used mainly for configuring the on-board ICs, while an Alveo U50 board (AMD/Xilinx) handles microwave-data transfer. The FPGA side interfaces through JESD204C to two AD9082 devices (Analog Devices, Inc.), each integrating four DACs~(12 GSPS, 16 bits) and two ADCs~(6 GSPS, 12 bits). To support high channel density while suppressing coupling between channels, the microwave paths are implemented with differential routing, baluns, and additional shielding; this arrangement allows 12 outputs and 4 inputs to be integrated on one board with compact spacing.

\begin{figure*}[t]
    \centering
    \begin{minipage}{0.95\textwidth}
        \raggedright
        {\sffamily\fontsize{9}{10.8}\selectfont\bfseries (a)}
        \par\vspace{0.2ex}
        \includegraphics[width=\linewidth]{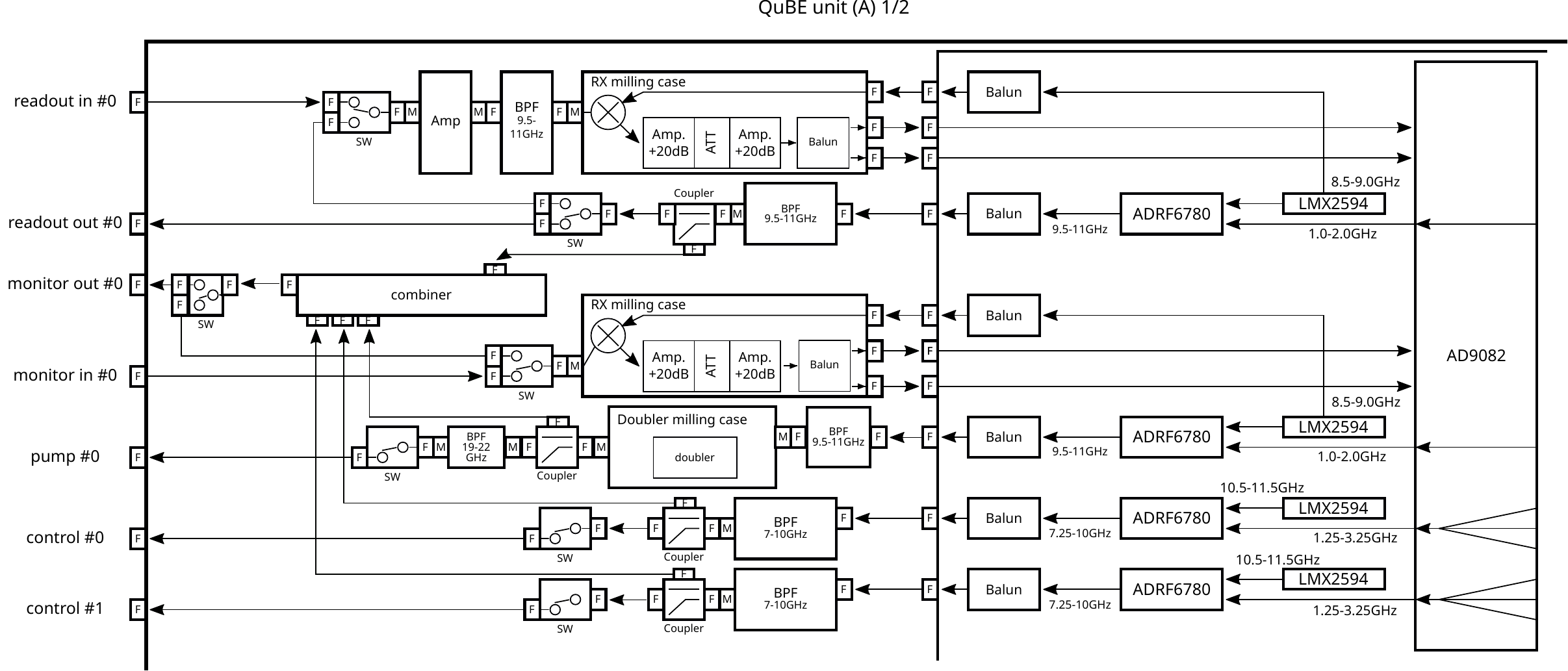}
        \par\vspace{1.0ex}
        {\sffamily\fontsize{9}{10.8}\selectfont\bfseries (b)}
        \par\vspace{0.2ex}
        \includegraphics[width=\linewidth]{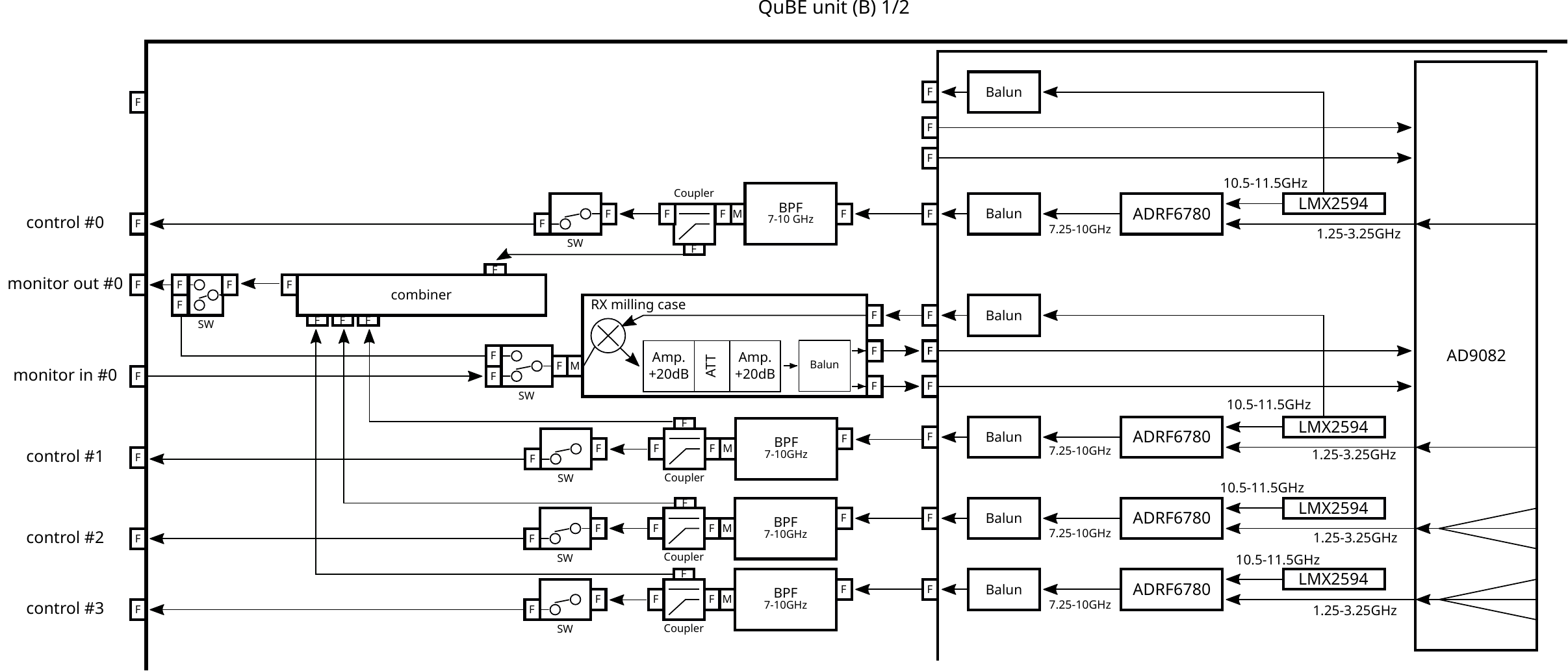}
    \end{minipage}
    \caption{\label{fig:qube-microwave-components} Detailed microwave-component layouts in the QuBE half units. (a) Type-A half unit, including readout receive, readout output, monitor, pump, and control paths. (b) Type-B half unit, including control and monitor paths. The diagrams show the switches, couplers, filters, baluns, mixers, PLL synthesizers, and AD9082 connections used to realize the signal paths summarized in Fig.~\ref{fig:hardware-architecture}(a).}
\end{figure*}

Signal paths per role are organized as follows, with the detailed microwave-component layouts of the two half-unit types shown in Fig.~\ref{fig:qube-microwave-components}. The control path generates arbitrary waveforms in the \qtyrangehy{1.25}{3.25}{\giga\hertz} range and mixes them with \qtyrangehy{10.5}{11.5}{\giga\hertz} local oscillators (generated by LMX2594 PLL synthesizers), with the lower sideband of the ADRF6780 mixer output selected to produce qubit-drive signals in the \qtyrangehy{7.25}{10}{\giga\hertz} band. Within the AD9082 chain, waveform data are distributed across digital channels, shifted first by FNCO paths and then by the CNCO, and finally converted to analog microwave output. In this implementation, three digital channels are combined for all control ports of Unit A and for four of the control ports of Unit B, yielding an instantaneous bandwidth of \qty{1.6}{\giga\hertz} on those outputs; the remaining Unit B control ports use a single digital channel. On the readout side, \qtyrangehy{1}{2}{\giga\hertz} waveform signals are mixed with \qtyrangehy{8.5}{9}{\giga\hertz} local oscillators, and the upper sideband is selected to generate \qtyrangehy{9.5}{11}{\giga\hertz} readout tones, while the receiver chain amplifies and downconverts the returning signals before differential delivery to the ADCs. Pump signals are generated through the same readout-style upconversion stage to produce \qtyrangehy{9.5}{11}{\giga\hertz} tones, which are then passed through an HMC814 frequency doubler so that \qtyrangehy{19}{21}{\giga\hertz} tones can be generated for JPA operation.

\begin{figure*}[t]
    \centering
    \includegraphics[width=0.95\textwidth]{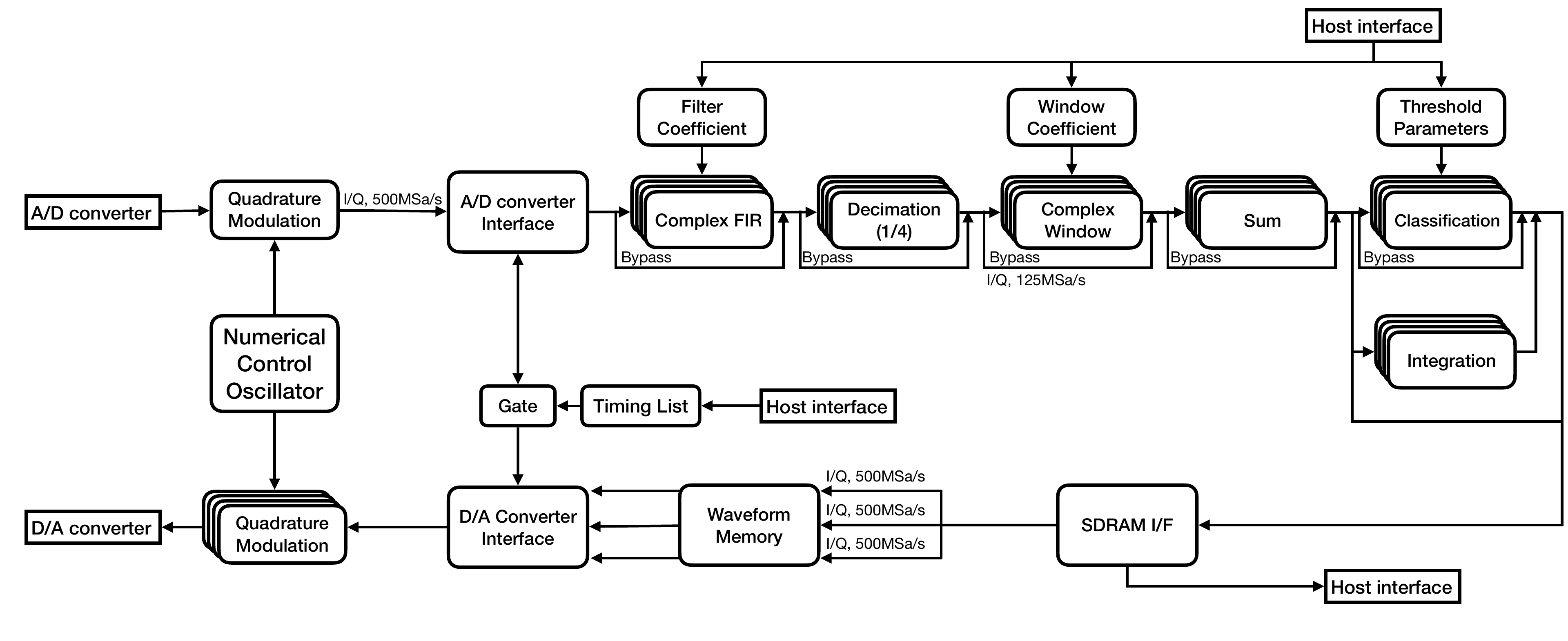}
    \caption{\label{fig:qube-fpga-dsp} Digital signal processing implemented in the FPGA for readout capture. The captured signal is digitally downconverted and demultiplexed into frequency components, filtered, decimated, multiplied by integration windows, and reduced to accumulated traces, integrated values, or threshold-classified outcomes for return to the host software.}
\end{figure*}

The unit also includes an internal monitoring path that supports in-situ pulse inspection and inter-unit timing checks. Solid-state single-pole double-throw~(SPDT) switches on the readout output and input ports allow the readout output to be internally looped back to the readout input, and directional couplers with a nominal coupling factor of \qty{-30}{\decibel} on the control, readout, and pump output paths route a fraction of selected signals into a shared downconversion and digitization stage. Unit B combines signals from four control ports, whereas Unit A combines signals from two control ports, one readout output port, and one pump output port. Isolation from the cryogenic wiring during monitoring is maintained at no less than \qty{60}{\decibel} by the SPDT switches together with an additional solid-state single-pole single-throw~(SPST) switch on each output channel. The monitor path routes the monitored signals through downconversion and digitization for waveform and timing inspection. The readout receiver chain provides the full DSP path for measurement capture, as summarized in Fig.~\ref{fig:qube-fpga-dsp}: CNCO/FNCO-based digital downconversion is followed by quadrature demodulation implemented in the FPGA, demultiplexing over up to four carrier frequencies, digital filtering, decimation, windowing, integration, and threshold-based classification into up to four categories. The processed readout data can be accumulated and returned to the host software as I/Q traces, integrated values, or classified outcomes. To reduce slow drift during long experiments, the main active microwave components are temperature stabilized using thermistors and Peltier elements attached to the AD9082, ADRF6780, LMX2594, and HMC814 devices, with feedback control implemented on the unit power board.

\section{Software details}
\label{sec:appendix-software}
\label{sec:appendix-software-details}

Within the stack, Qubex is the experiment-facing framework whose system, pulse, measurement, and experiment layers are described in Sec.~\ref{sec:architecture}. Qubex lets experiment routines be written in terms of logical targets, pulse schedules, and measurement options, while resolving them to the channels, frequencies, timing parameters, and calibration state needed for execution on the selected control system. The experiment layer provides workflows for procedures such as resonator and qubit spectroscopy, DRAG-based single-qubit calibration, coherence-time characterization, multiplexed single-shot readout with state classification, cross-resonance calibration with Hamiltonian tomography and active cancellation, and randomized benchmarking. Returned data and fitted quantities are exposed through Qubex result objects rather than through backend-specific payloads.

\begin{figure*}[t]
    \centering
    \includegraphics[width=\textwidth]{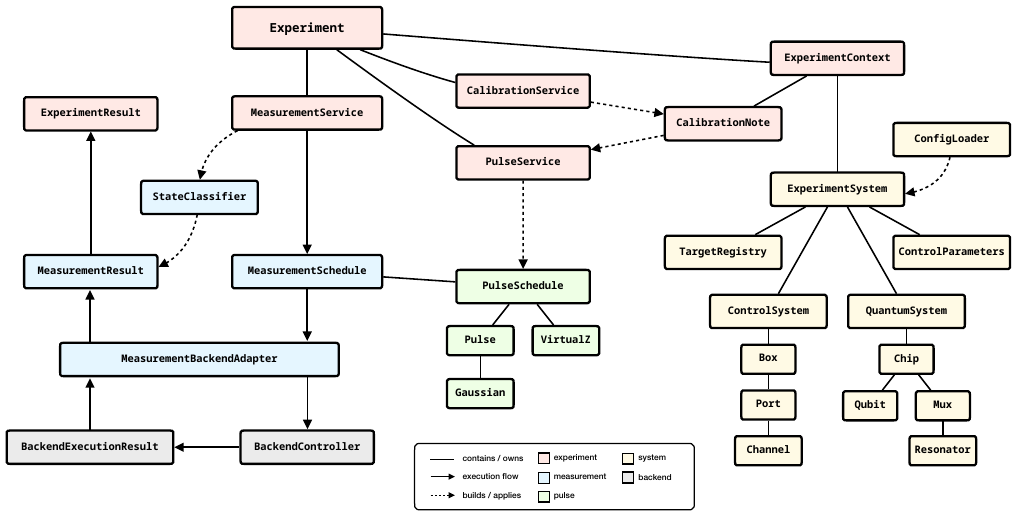}
    \caption{\label{fig:qubex-software-details} Implementation-level organization of Qubex v1.5.0 objects used during experiment execution. Box colors indicate object layers or responsibility groups: experiment, measurement, pulse, system, and backend. Solid lines indicate ownership or containment, solid arrows indicate execution flow, and dashed arrows indicate relationships in which one object builds or applies another.}
\end{figure*}

Fig.~\ref{fig:qubex-software-details} provides an object-level view of how Qubex represents experiment workflows, system state, pulse schedules, backend execution, and returned results. At initialization, \xCode{ConfigLoader} reads configuration files and builds a setup-specific \xCode{ExperimentSystem}. This setup-resolution step assigns controller channels, sets the LO, CNCO, and FNCO terms used by those channels, determines the residual AWG-side modulation frequency from the target and fine-frequency settings, loads clock and skew-related settings, and connects the target registry and control parameters to the modeled control and quantum systems. The \xCode{ControlSystem} side represents hardware resources such as boxes, ports, and channels, whereas the \xCode{QuantumSystem} side represents chip objects such as qubits, readout-multiplexing units, and resonators. Using this resolved model, the \xCode{TargetRegistry} maps experiment-level target labels to microwave-control targets with assigned controller channels and frequency settings, including qubit \(g\)--\(e\) and \(e\)--\(f\) transitions, readout-resonator targets, and CR-drive targets for cross-resonance tones. \xCode{ControlParameters} stores setup-specific experimental settings such as drive amplitudes, capture delays, and JPA-pump settings. Together, these objects provide the state against which experiment routines resolve logical targets into executable controller settings.

Users interact with the \xCode{Experiment} API rather than with hardware commands directly. The \xCode{Experiment} object owns the \xCode{ExperimentContext}, which carries the resolved setup state and run-time information, and delegates concrete tasks to service objects. \xCode{MeasurementService} constructs and dispatches measurement definitions. \xCode{CalibrationService} executes calibration routines, including single- and two-qubit gate calibrations, and writes the resulting parameters to \xCode{CalibrationNote}. \xCode{PulseService} then uses these parameters to assemble the concrete pulse sequences required by each experiment. Because the updated \xCode{CalibrationNote} is part of the \xCode{ExperimentContext}, subsequent pulse construction and measurement execution resolve their settings from the refreshed calibration state.

The pulse representation is centered on \xCode{PulseSchedule}. A schedule contains waveform objects, blanks, channel labels, timing barriers, frequency metadata, and virtual-\(Z\) frame-shift operations, and it can be composed hierarchically by inserting one schedule into another. The inserted schedule is aligned on the participating channels and remains symbolic until waveform samples are required. At sampling or backend lowering time, the hierarchy is resolved into per-channel waveform sequences; accumulated virtual-\(Z\) shifts are applied as phase updates to subsequent pulses, and any remaining frame shift is retained as the final frame state. This allows composite gates and calibration routines to be assembled from smaller schedule blocks without manually resampling intermediate pulses or rewriting phase corrections. Because \xCode{PulseSchedule} is a control-system-independent representation rather than a controller-register payload, the same pulse-construction logic can be shared across different control-hardware and simulator paths while still being lowered later to backend-specific instructions.

The measurement layer turns a pulse schedule into an execution-ready measurement request by combining it with readout and capture information. For one measurement point, \xCode{MeasurementSchedule} contains the pulse and capture schedules: readout pulses and capture windows are placed in a common time frame, and backend constraints are applied during construction and validation. Runtime acquisition options such as shot count, averaging, integration, and classification are supplied separately through the measurement configuration at execution time. Parameter sweeps are handled above this object boundary: each sweep point is represented by a pointwise measurement request, while the sweep values, axis order, and array shape are stored in the returned sweep-result objects. This keeps \xCode{MeasurementSchedule} focused on a single executable measurement request while preserving the association between varied parameters, logical targets, and returned data.

Backend-specific adapters are used at the boundary between \xCode{MeasurementSchedule} and \xCode{BackendController}. The adapter selected for the active controller validates backend constraints and converts the common schedule and measurement configuration into a backend-specific payload. In the hardware backend used for the QuBE experiments reported here, this payload contains sampled generation and capture sequences, resource maps, repetition and averaging options, and capture-processing flags. The backend controller then hands the payload to the lower-level qubecalib/quelware stack, which performs hardware transport, register configuration, execution, and readout. For multi-unit experiments, the backend builds the waveform-generation and capture actions of all participating units, triggers their execution at the scheduled counter value, and collects the capture results returned by each unit into a single payload.

Returned backend payloads are normalized before they are used by experiment routines, so that experiment code sees one target-keyed data product regardless of how many units produced the underlying captures. The \xCode{MeasurementResult} object stores target-keyed capture data together with the measurement configuration and optional classifier references; the capture payloads can contain waveform traces, integrated I/Q values, or classified state series depending on the acquisition mode. For parameter sweeps, the surrounding sweep-result objects keep the sweep values, axis order, and array shape while holding the pointwise \xCode{MeasurementResult} objects. \xCode{StateClassifier} objects can then be applied to the normalized readout data to assign measurement outcomes, including multilevel classifications. When Qubex-side classification is used, \xCode{StateClassifier} can apply scikit-learn-based Gaussian-mixture-model or \(k\)-means classifiers~\cite{pedregosa2011scikit}. Higher-level workflows package the relevant measurement results, classified outcomes, and fitted quantities into \xCode{ExperimentResult} objects, which provide protocol-specific data-analysis and visualization interfaces for the corresponding experiment.

The object organization described above defines the Qubex side of the software boundary. Qubex retains the experiment model, control-system-independent pulse representation, measurement construction, and result-analysis contracts. Beneath the Qubex backend boundary, qubecalib provides intermediate support for functions such as multiplexed measurement handling and skew adjustment, bridging the experiment-facing interface and the hardware runtime. At the hardware-control layer used for the QuBE system, quelware performs direct hardware control, with its central package \xCode{quel\_ic\_config} handling register-level IC configuration through the exStickGE path, and e7awgsw controls waveform transfer, execution, and capture-side digital signal processing on the Alveo U50 path. This division keeps hardware-specific transport and register configuration below the backend boundary while allowing experiment-facing workflows to remain written in terms of logical targets, schedules, calibration state, and returned data products.

The resulting control environment thus comprises Qubex together with the lower-level components that configure, execute, and read out the hardware stack. Fig.~\ref{fig:software-stack-versions} lists the software packages used in this stack, their versions, and the corresponding open-source repositories.

\begin{figure}[t]
    \centering
    \begin{ruledtabular}
        \begin{tabular*}{\columnwidth}{@{\extracolsep{\fill}}lll}
            Software  & Version & Repository            \\
            \colrule
            Qubex     & v1.5.0  & \href{https://github.com/amachino/qubex}{amachino/qubex}        \\
            qubecalib & v3.1.15 & \href{https://github.com/qiqb-osaka/qube-calib}{qiqb-osaka/qube-calib} \\
            quelware  & v0.10.8 & \href{https://github.com/quel-inc/quelware}{quel-inc/quelware}     \\
            e7awgsw   & v1.0.0  & \href{https://github.com/e-trees/e7awg_sw}{e-trees/e7awg\_sw}     \\
        \end{tabular*}
    \end{ruledtabular}
    \caption{\label{fig:software-stack-versions} Software packages used in the QuBE/Qubex control stack. Repository entries identify the corresponding GitHub repositories.}
\end{figure}
\section{Experimental details}
\label{sec:appendix-experimental-details}

\subsection{Cryostat wiring}
\label{sec:appendix-cryostat-wiring}

\begin{figure}[t]
    \centering
    \includegraphics[width=1.0\columnwidth]{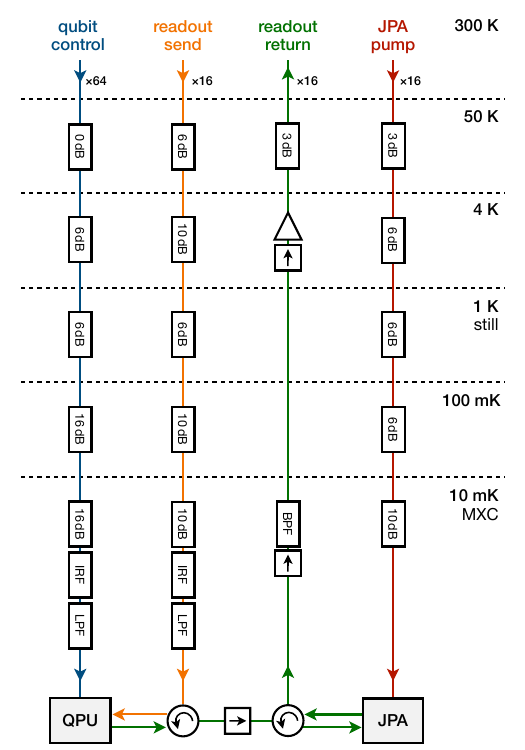}
    \caption{\label{fig:cryostat-wiring} Cryostat wiring for the 64-qubit experiment. Representative attenuation and filtering stages are distributed across the \qtyhy{50}{\kelvin}, \qtyhy{4}{\kelvin}, still (\qtyhy{1}{\kelvin}), \qtyhy{100}{\milli\kelvin}, and MXC (\qtyhy{10}{\milli\kelvin}) stages to suppress thermal noise from higher-temperature stages while spreading the associated heat load. The readout-return path includes circulators (circular-arrow), isolators (square-arrow), a band-pass filter~(BPF), and a HEMT amplifier. IRF and LPF denote infrared and low-pass filters, respectively.}
\end{figure}

Fig.~\ref{fig:cryostat-wiring} summarizes the cryostat wiring used in the 64-qubit experiment. The cryostat inputs consist of qubit-control, readout-send, and JPA-pump lines, and the cryostat output is the readout-return line. Each qubit has a dedicated control line, whereas the readout-send, readout-return, and JPA-pump paths are shared within each four-qubit readout-multiplexing unit. This line assignment gives 64 control lines and 16 lines of each shared type.

The chip is mounted at the mixing-chamber stage~(MXC) of the dilution refrigerator and housed inside a magnetic shield to suppress stray-field-induced flux noise. Each control line passes through attenuation and filtering stages at the \qtyhy{50}{\kelvin}, \qtyhy{4}{\kelvin}, still (\qtyhy{1}{\kelvin}), \qtyhy{100}{\milli\kelvin}, and MXC (\qtyhy{10}{\milli\kelvin}) temperature stages before reaching the qubit chip. These stages suppress thermal noise from higher-temperature stages and limit out-of-band microwave components.

Within each readout-multiplexing unit, the readout-send line delivers frequency-multiplexed readout tones to the readout resonators, and the readout-return line carries the reflected signals back through the circulator and isolators. The JPA-pump line supplies the pump tone to the ImPA, which provides the first low-noise amplification stage for superconducting qubit readout~\cite{yamamoto2008flux,lin2013single}. The HEMT amplifier at the \qtyhy{4}{\kelvin} stage then provides further low-noise cryogenic amplification before the signal returns to the QuBE readout input ports for FPGA-based digitization and demodulation.

\subsection{Resolved hardware frequency settings}
\label{sec:appendix-frequency-settings}

Qubex resolves each logical target frequency into a hardware frequency setting that satisfies the discrete grids and bandwidth limits of the QuBE microwave path. The decomposition follows Eq.~\eqref{eq:hardware-frequency-decomposition}, with the LO, CNCO, FNCO, and AWG terms chosen so that the generated tone remains phase reproducible and avoids unwanted aliasing or image tones. The LO is constrained to a \qtyhy{500}{\mega\hertz} grid so that the LO phase is unchanged at the \qtyhy{2}{\nano\second} waveform-sampling boundary. The CNCO and FNCO are constrained to the NCO grid used by the backend frequency-tuning words and FPGA block timing; in the present implementation this grid is \qty{23.4375}{\mega\hertz}, corresponding to three cycles over the \qtyhy{128}{\nano\second} minimum block. The remaining residual frequency \(F_{\mathrm{AWG}}\) is left to the sampled pulse waveform and is kept small enough to avoid aliasing and waveform-distortion artifacts.

The decomposition is not performed independently for each target. A single physical output port must often cover several logical targets, such as the qubit \(g\)--\(e\) transition and the CR tones that address neighboring target-qubit frequencies. Qubex therefore first chooses a role-dependent CNCO center within the supported mixer/NCO operating range, then chooses a shared LO and CNCO for the port, and finally assigns per-channel FNCO offsets. For readout, the shared setting is centered on the resonator frequencies within one readout-multiplexing unit. For control, the setting depends on the available number of digital channels: in the configuration used here, one channel covers the resonant single-qubit drive and the remaining channels cover CR-drive target groups.

For the CR channels, Qubex collects the neighboring target-qubit frequencies that can be driven from a given control-qubit line and groups them by frequency. When two CR channels are available, the grouping is chosen to minimize the largest within-group frequency span, and the FNCO of each CR channel is placed near the corresponding group center. Individual CR targets within that group are then reached by the residual AWG-side modulation \(F_{\mathrm{AWG}}\). Qubex chooses these terms to avoid aliasing by keeping the AWG frequency within \(\pm\qty{200}{\mega\hertz}\) and the FNCO frequency within \(\pm\qty{600}{\mega\hertz}\); settings that cannot satisfy these margins are reported as errors or warnings. This strategy favors covering as many useful CR targets as possible while keeping each waveform modulation close to baseband.

Table~\ref{tab:control-frequency-settings} gives the resolved settings used for the Q28-centered control and readout targets. The table lists the target frequency, the LO frequency, the CNCO and FNCO components, the residual AWG-side frequency term deployed in the pulse waveform, and the selected sideband. The far-detuned CR benchmark in Sec.~\ref{sec:cross-resonance} uses the Q28--Q25 setting in this table.

\begin{table*}[t]
    \caption{\label{tab:control-frequency-settings} Resolved hardware frequency settings auto-generated by Qubex for a Q28-centered set of control and readout targets. Labels of the form Q28--Q25 denote CR-drive settings for a CR operation with Q28 as the control qubit and Q25 as the target qubit; RQ28 denotes the readout-frequency target for Q28. All frequencies are in megahertz.}
    \begin{ruledtabular}
        \begin{tabular*}{\textwidth}{@{\extracolsep{\fill}}crrrrrc}
            Label     & $F_{\mathrm{target}}$ & $F_{\mathrm{LO}}$ & $F_{\mathrm{CNCO}}$ & $F_{\mathrm{FNCO}}$ & $F_{\mathrm{AWG}}$ & Sideband \\
            \colrule
            Q28       & $7971.349$            & $10500$          & $2062.500$          & $+468.750$          & $-2.599$          & LSB      \\
            Q28--Q14  & $8859.447$            & $10500$          & $2062.500$          & $-468.750$          & $+46.803$         & LSB      \\
            Q28--Q25  & $8700.633$            & $10500$          & $2062.500$          & $-210.938$          & $-52.196$         & LSB      \\
            Q28--Q29  & $8904.411$            & $10500$          & $2062.500$          & $-468.750$          & $+1.839$          & LSB      \\
            Q28--Q30  & $8658.505$            & $10500$          & $2062.500$          & $-210.938$          & $-10.068$         & LSB      \\
            RQ28      & $10127.055$           & $8500$           & $1710.938$          & $+0.000$            & $-83.882$         & USB      \\
        \end{tabular*}
    \end{ruledtabular}
\end{table*}

\subsection{Cross-resonance model and calibration}
\label{sec:appendix-cr-details}

For the far-detuned CR gate used in Fig.~\ref{fig:cross-resonance}, the drive applied to the control qubit induces not only the desired conditional interaction on the target qubit but also unwanted single-qubit and crosstalk terms~\cite{chow2011simple, sheldon2016crosstalk}. In practice, the calibration therefore aims to isolate the useful conditional rotation while suppressing the unwanted rotations that appear on the target qubit and in the residual two-qubit terms.

Following the CR Hamiltonian tomography convention used for crosstalk calibration~\cite{sheldon2016crosstalk}, we describe the driven interaction in the two-qubit computational subspace as
\begin{align}
    H_{\mathrm{CR}}
     & =
    \frac{\hbar}{2}
    \left(
    \Omega_{IX} IX
    + \Omega_{IY} IY
    + \Omega_{IZ} IZ
    \right.
    \nonumber      \\
     & \quad\left.
    + \Omega_{ZX} ZX
    + \Omega_{ZY} ZY
    + \Omega_{ZZ} ZZ
    \right).
\end{align}
Here, each term \(\Omega_{PQ}PQ\) represents an effective rotation-rate contribution for the two-qubit Pauli product \(PQ\), where \(P\) acts on the control qubit and \(Q\) acts on the target qubit. The desired conditional interaction is the \(ZX\) component, while \(IX\), \(IY\), \(IZ\), \(ZY\), and \(ZZ\) represent unwanted target rotations and residual two-qubit terms. In this convention the target-side cancellation tone is not written as a separate Hamiltonian term; instead, its amplitude and phase are tuned so that the measured \(IX\) and \(IY\) coefficients are suppressed while the calibrated \(ZX\) rate is preserved.

In Qubex, the effective interaction is characterized by CR Hamiltonian tomography. For a selected control--target pair, Qubex prepares the control qubit in $\ket{g}$ and $\ket{e}$, applies CR drives of varied duration, and performs state tomography on the resulting target-qubit trajectory. The target Bloch-vector trajectories are fitted to a damped constant-axis rotation model: Qubex evaluates the coherent rotation using Rodrigues' rotation formula and multiplies it by a fitted exponential decay envelope. The two fitted rotation generators for the control qubit prepared in \(\ket{g}\) and \(\ket{e}\) give the average and conditional rotation components. The average part is interpreted as control-state-independent target rotation, whereas the difference isolates the conditional terms. In the notation above, this procedure extracts the \(IX\), \(IY\), \(IZ\), \(ZX\), \(ZY\), and \(ZZ\) coefficients used to separate the desired conditional \(ZX\) rotation from unwanted target rotations.

The first calibration update uses these tomography coefficients to align the effective CR rotation axis and to set the active cancellation tone. The phase of the measured complex coefficient \(\Omega_{ZX}+i\Omega_{ZY}\) determines the phase correction applied to the CR pulse so that the useful conditional response is aligned with the intended \(ZX\) axis. The measured complex coefficient \(\Omega_{IX}+i\Omega_{IY}\) determines the amplitude and phase of a target-side cancellation pulse that is added with the opposite phase to suppress the unwanted target rotation. The calibration result records the updated CR amplitude and phase, cancellation amplitude and phase, ramp time, and measured \(\Omega_{ZX}\), so that later pulse construction starts from tomography-derived quantities rather than from manually specified waveform settings.

In the multi-unit configuration of Fig.~\ref{fig:cross-resonance}(a), the CR pulse at \(f_{\mathrm{t}}\) and the target-side cancellation and rotary pulses at the same frequency are generated by different QuBE units with independent PLL-based LOs. Phase coherence between these tones is maintained by construction: all LOs are phase-locked to the shared \qtyhy{100}{\mega\hertz} reference, and the LO, NCO, and AWG frequency terms are chosen on the grids described in Appendix~\ref{sec:appendix-frequency-settings} so that the phase of each output is reproducible at the scheduled pulse boundaries. The remaining static phase offset between the two output paths is not tracked separately; it is absorbed into the calibrated CR and cancellation phases extracted by the tomography-based calibration, which are stored in the Qubex calibration state and reused by subsequent pulse construction. Slow drift of the relative phase is expected to remain at the scale of the output-phase stability measured in Sec.~\ref{sec:hardware-characterization}, with standard deviations below \qty{0.456}{\degree} over \qty{3600}{\second} across the measured ports.

The measured \(\Omega_{ZX}\) provides an initial estimate for a \(ZX_{\pi/2}\) pulse, which is then calibrated using an echoed sequence. A single CR block consists of a shaped CR pulse on the control-qubit drive channel and a simultaneous shaped cancellation pulse on the target-qubit channel. The echoed sequence applies one CR block, a control-qubit $\pi$ pulse, a second CR block with the CR drive sign inverted, and a second control-qubit $\pi$ pulse. This construction preserves the sign-changing conditional response while canceling rotations that do not reverse with the CR drive. The gate calibration then sweeps the CR amplitude using this echoed construction, compares responses from repeated sequences to locate the \(ZX_{\pi/2}\) condition, and updates the stored gate duration and drive amplitude. The target-side pulse used during the final gate is the vector sum of the cancellation component and a rotary component. This final pulse-construction step is where the \(IX\)-like target rotation used for the rotary echo is intentionally included: the cancellation component suppresses the tomography-measured \(\Omega_{IX}+i\Omega_{IY}\) component, while the added rotary component is set from the measured \(\Omega_{ZX}\) to reduce residual coherent error during the echoed gate~\cite{sundaresan2020rotary}.

The Q28--Q25 gate used for this benchmark was calibrated by running the built-in Qubex two-qubit calibration procedure with its default settings. Table~\ref{tab:cr-gate-parameters} summarizes the calibrated parameters used for the CR benchmark in Fig.~\ref{fig:cross-resonance}(b). Drive amplitudes are given as Rabi-equivalent amplitudes; the CR-drive conversion uses the Q28 calibration, and the cancellation and rotary conversions use the Q25 calibration. The two-qubit Clifford sequences were compiled from the calibrated \(ZX_{\pi/2}\) gate, single-qubit \(X_{\pi/2}\) pulses, and virtual-\(Z\) frame updates. For the IRB measurement, each data point averaged 30 random Clifford sequences, and each sequence was measured with 4096 shots.

As a reference scale, we estimate the coherence-only average fidelity using the coherence-limit expression of Ref.~\cite{wei2024characterizing}, which assumes independent amplitude-damping and dephasing channels during the gate duration. Using the \(T_1\) and \(T_{2,\mathrm{echo}}\) values in Table~\ref{tab:device-parameters} and the echoed-gate duration of \(\qty{240}{\nano\second}\) in Table~\ref{tab:cr-gate-parameters} gives \(F_{\mathrm{coh}}\simeq 99.13\%\). The measured IRB fidelity of \(98.34\%\) is therefore about \(0.8\) percentage points below this estimate, indicating that the calibrated gate is close to the estimated coherence-limited fidelity while leaving room for improvements in residual control, crosstalk, and calibration errors.

\begin{table}[t]
    \caption{\label{tab:cr-gate-parameters} Pulse parameters for the calibrated CR benchmark in Fig.~\ref{fig:cross-resonance}(b).}
    \begin{ruledtabular}
        \begin{tabular*}{\columnwidth}{@{\extracolsep{\fill}}lc}
            Parameter                      & Value                                      \\
            \colrule
            Control--target pair           & Q28--Q25                          \\
            Control-qubit frequency \(f_{\mathrm{c}}\) & \(\qty{7971.349}{\mega\hertz}\)   \\
            Target-qubit frequency \(f_{\mathrm{t}}\) & \(\qty{8700.633}{\mega\hertz}\)   \\
            Qubit detuning \(\Delta_{\mathrm{ct}}=f_{\mathrm{c}}-f_{\mathrm{t}}\) & \(\qty{-729.284}{\mega\hertz}\) \\
            CR-pulse duration              & \(\qty{96.0}{\nano\second}\)      \\
            CR-pulse ramp duration         & \(\qty{16.0}{\nano\second}\)      \\
            Control \(\pi\)-pulse duration & \(\qty{24.0}{\nano\second}\)      \\
            Echoed-gate duration           & \(\qty{240.0}{\nano\second}\)     \\
            CR-drive Rabi amplitude \(\Omega_{\mathrm{R},\mathrm{c}}/2\pi\) & \(\qty{322.8}{\mega\hertz}\) \\
            CR-drive phase \(\phi_{\mathrm{CR}}\) & \(\qty{2.920}{\radian}\) \\
            Cancellation Rabi amplitude \(\Omega_{\mathrm{R},\mathrm{t}}^{\mathrm{cancel}}/2\pi\) & \(\qty{7.9}{\mega\hertz}\) \\
            Cancellation phase \(\phi_{\mathrm{cancel}}\) & \(\qty{-3.097}{\radian}\) \\
            Rotary Rabi amplitude \(\Omega_{\mathrm{R},\mathrm{t}}^{\mathrm{rot}}/2\pi\) & \(\qty{11.1}{\mega\hertz}\) \\
        \end{tabular*}
    \end{ruledtabular}
\end{table}

\subsection{Multilevel-readout assignment and error analysis}
\label{sec:appendix-multilevel-readout}

To construct the qubit-state classifier, Qubex prepares $\ket{g}$, $\ket{e}$, and $\ket{f}$ separately for the measured qubit and records the integrated complex I/Q values for each prepared state. The data in Fig.~\ref{fig:multilevel-readout} were acquired with a readout pulse of \qty{1.024}{\micro\second}. For each pump condition, 5000 shots were acquired for each prepared state. The I/Q samples from the three prepared states are pooled so that a single three-component Gaussian-mixture model captures the three readout clusters corresponding to $\ket{g}$, $\ket{e}$, and $\ket{f}$. The fitted Gaussian components are then labeled as $\ket{g}$, $\ket{e}$, or $\ket{f}$ according to the prepared-state data, defining the classifier. The resulting classifier converts integrated I/Q shots into discrete state labels.

Table~\ref{tab:multilevel-readout-assignment-matrix} shows the assignment matrix for the JPA-pump-enabled multilevel-readout data in Sec.~\ref{sec:multilevel-readout}. Rows indicate the prepared transmon state, and columns indicate the assigned state from the trained Gaussian-mixture classifier. The diagonal elements give the state-assignment fidelities used in the main text. The assignment matrix is evaluated on the same dataset used to train the classifier; no separate held-out dataset was used. The largest off-diagonal element is the assignment of shots prepared in $\ket{f}$ as $\ket{e}$, corresponding to \(372/5000=7.44\%\). The next-largest off-diagonal element is the assignment of shots prepared in $\ket{e}$ as $\ket{g}$, corresponding to \(200/5000=4.00\%\).

\begin{table}[t]
    \caption{\label{tab:multilevel-readout-assignment-matrix} Assignment matrix for the JPA-pump-enabled multilevel-readout measurement on Q10. Entries show counts, with row-normalized assignment probabilities in parentheses.}
    \begin{ruledtabular}
        \begin{tabular*}{\columnwidth}{@{\extracolsep{\fill}}c r@{\,}l r@{\,}l r@{\,}l}
            Prepared & \multicolumn{2}{c}{$\ket{g}$} & \multicolumn{2}{c}{$\ket{e}$} & \multicolumn{2}{c}{$\ket{f}$} \\
            \colrule
            $\ket{g}$      & 4973 & (99.46\%) & 14   & (0.28\%)  & 13   & (0.26\%)  \\
            $\ket{e}$      & 200  & (4.00\%)  & 4784 & (95.68\%) & 16   & (0.32\%)  \\
            $\ket{f}$      & 68   & (1.36\%)  & 372  & (7.44\%)  & 4560 & (91.20\%) \\
        \end{tabular*}
    \end{ruledtabular}
\end{table}

The two largest off-diagonal elements follow the relaxation direction in the transmon ladder, $\ket{f}\rightarrow\ket{e}$ and $\ket{e}\rightarrow\ket{g}$. We therefore interpret the dominant assignment errors as primarily relaxation-related. However, this assignment matrix alone does not determine the underlying transition rates, because the off-diagonal counts can also include state-preparation error, readout-induced transitions, finite classifier overlap, and thermal population present before the prepared pulse sequence.

\begin{acknowledgments}
    The authors thank András Gunyhó for helpful feedback on the manuscript.

    This work was supported in part by JST COI-NEXT (Grant No.~JPMJPF2014), JST Moonshot R\&D (Grant Nos.~JPMJMS2067 and JPMJMS256F), MEXT Q-LEAP (Grant Nos.~JPMXS0118068682 and JPMXS0120319794), and JST, the establishment of university fellowships towards the creation of science technology innovation and JST SPRING (Grant No.~JPMJSP2138).

\end{acknowledgments}

\xBackmatterSection{Conflict of interest}

Some authors are inventors on related Japanese Patent No.~JP7303506B2, entitled ``Quantum computer control device.'' Some authors are affiliated with QuEL, Inc., which commercializes technology transferred from the University of Osaka in connection with the QuBE control system. The authors declare no other competing interests.

\xBackmatterSection{Data and code availability}

Figure/table data, metadata, and scripts are archived under DOI:~\href{https://doi.org/10.5281/zenodo.20625114}{10.5281/zenodo.20625114}; the Qubex software archive is available under DOI:~\href{https://doi.org/10.5281/zenodo.20621646}{10.5281/zenodo.20621646}.

\end{document}